# Recent progress of scanning tunneling microscopy/spectroscopy study of pair density wave in superconductors

Zi-Ang Wang, Bin Hu, Xianghe Han, Hui Chen*, and Hong-Jun Gao*

*Beijing National Center for Condensed Matter Physics and Institute of Physics, Chinese Academy of Sciences, Beijing 100190, PR China*

*Address correspondence to

Hui Chen, hchenn04@iphy.ac.cn; Hong-Jun Gao, hjgao@iphy.ac.cn

**Abstract**

A pair density wave (PDW) is a superconducting state characterized by an order parameter with finite center-of-mass momentum in the absence of an external magnetic field, thereby breaking the conventional translational symmetry in homogeneous superconductors. It is proposed that PDW emerges from magnetic interactions, strong electron-electron correlations, and their interplay with competing orders. In this review, we highlight recent advances in the detection and study of PDWs using scanning tunneling microscopy and spectroscopy (STM/STS). We focus on how the signatures of PDW have been experimentally visualized across a variety of extraordinary superconductors, including iron-based superconductors, cuprate superconductors, spin-triplet superconductors, kagome-lattice superconductors, and transition metal dichalcogenides. Beginning with an introduction to the fundamental concept of PDWs and the unique capabilities of STM/STS, particularly its atomic-scale spatial resolution and advanced data analysis techniques, we discuss key experimental findings, including the direct visualization of charge density modulations associated with PDWs. Finally, we discuss emerging challenges and future directions, aiming to inspire future research into the nature of PDWs in superconductors.

## 1. Introduction

### 1.1 Finite momentum Cooper pairs and pair density wave

The exploration of unconventional superconducting states has revealed a wide range of phenomena that extend beyond the predictions of the Bardeen-Cooper-Schrieffer (BCS) theory, which describes spin-singlet superconducting (SC) states dominated by Cooper pairs with zero center-of-mass momentum [1]. In this conventional framework, the momenta and spins of the paired electrons are opposite, resulting in a spatially uniform superconducting order parameter (Figure 1(a,b)). In some particular cases, SC states with non-zero center-of-mass momentum Cooper pairs have been proposed. One notable example is the Fulde-Ferrell-Larkin-Ovchinnikov (FFLO) state, which emerges under specific conditions where the superconducting order parameter exhibits a spatial modulation [2, 3]. First proposed in the 1960s, the FFLO state appears in superconductors subjected to strong external magnetic fields, where the Zeeman effect disrupts the balance between spin-up and spin-down Fermi surfaces and breaks the symmetry of Cooper pairing. In contrast to conventional Cooper pairs that form with zero center-of-mass momentum and result in a translation-invariant superconducting order parameter (Figure 1(a,b)) [4], the electron pairs in the FFLO state acquire finite momentum $\boldsymbol{Q}$ because of the momentum mismatch between spin-up and spin-down Fermi surfaces, as illustrated in Figure 1(c,d). In this scenario, two cases have been proposed: (i) the Fulde-Ferrell (FF) state, in which the order parameter takes the form $\Delta(\boldsymbol{r}) \propto e^{i\boldsymbol{Q}\cdot\boldsymbol{r}}$ and exhibits a spatially varying phase while retaining a uniform amplitude; and (ii) the Larkin-Ovchinnikov (LO) state, in which the order parameter $\Delta(\boldsymbol{r}) \propto \cos(\boldsymbol{Q}\cdot\boldsymbol{r})$ exhibits a real-space amplitude modulation. In general, such FFLO states possess a spatially modulated order parameter whose spatial average vanishes.

The FFLO state can be regarded as an early example of a finite-momentum pairing state [5]. In a broader sense, such finite-momentum pairing states can be categorized under the general concept of PDWs, with the FFLO state being one realization under an external magnetic field. A PDW order is characterized by Cooper pairs carrying finite center-of-mass momentum $\boldsymbol{Q}_{\mathrm{pdw}}$, which breaks the lattice translational symmetry (Figure 1(e,f)). As a result, the SC order parameter exhibits a periodic modulation in real space from one unit cell of PDW to another [6–8]. It is worth noting that, in a strict sense, a pure PDW refers to a SC state in which the order parameter exhibits spatial modulation without a uniform component [9–11], i.e., $\Delta(\boldsymbol{r}) = \Delta_{\mathrm{P}}\cos(\boldsymbol{P}\cdot\boldsymbol{r})$, where $\Delta_{\mathrm{P}}$ denotes the modulation amplitude and $\boldsymbol{P}$ the corresponding wavevector. In contrast, a spatially uniform superconductor is characterized by a finite uniform component ($\Delta_{\boldsymbol{P}=0} \neq 0$ and $\Delta_{\boldsymbol{P}\neq0} = 0$) [9, 11]. While experimentally, finite Fourier components of $\Delta(\boldsymbol{r})$ at reciprocal lattice vectors may be observed which reflect the underlying lattice periodicity rather than a broken translational symmetry. Therefore, in the strict theoretical definition, a pure PDW is a distinct quantum phase of matter, characterized by $\Delta_{\boldsymbol{P}=0} = 0$ and $\Delta_{\mathrm{P}\neq0} \neq 0$, which breaks the lattice translational symmetry [11]. Many theoretical studies support this definition; however, in almost all experimental situations, uniform superconductivity coexists with modulated components, i.e., $\Delta(\boldsymbol{r}) = \Delta_0 + \Delta_{\mathrm{P}}\cos(\boldsymbol{P}\cdot\boldsymbol{r})$, where $\Delta_0$ denotes the uniform component. In such cases, the observed modulations are often interpreted as signatures of a PDW intertwined with uniform superconductivity. Furthermore, PDW states often coexist or compete with other ordered phases, such as the uniform SC states, charge density waves (CDWs) and spin density waves (SDWs) [12, 13], which will be discussed later in this introduction.

The study of PDW holds broad and significant implications, particularly for the field of topological superconductivity [14, 15]. PDWs are believed to provide a natural platform for hosting exotic quasiparticles such as Majorana bound states, which are of great interest for quantum computation [16–19]. Beyond this, PDWs are believed to play a key role in unraveling the mysterious pseudogap phase observed in cuprates [6]. In systems where PDWs coexist with or compete against other ordered phases, novel quantum states of matter may emerge [6, 12], including charge 4*e* or 6*e* superconducting states [20–22]. These emergent phenomena not only deepen the understanding of superconducting pairing mechanisms but also open new avenues for exploring the complex landscape of strongly correlated electronic systems.

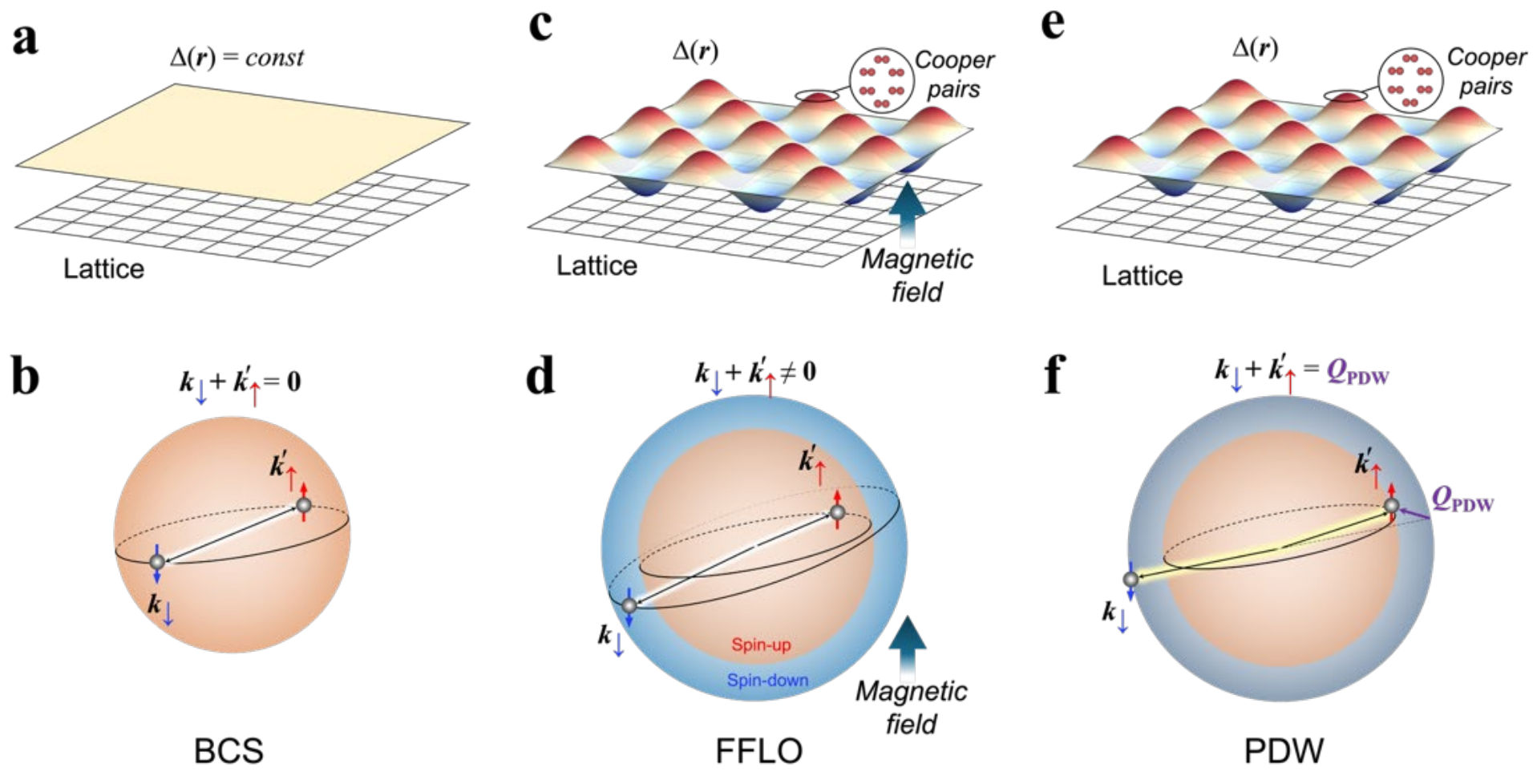

**Fig. 1** Schematic illustration of conventional BCS superconductivity, FFLO states, and PDW states. (**a**, **b**) In a conventional BCS superconductor, electrons with opposite spins and momenta form Cooper pairs, resulting in a zero center-of-mass momentum and a spatially uniform SC order parameter. (**c**, **d**) In the FFLO state, an external magnetic field results in the imbalance between spin-up and spin-down Fermi surfaces, leading to a finite pairing momentum and a spatially modulated order parameter. (**e**, **f**) In the PDW state, a non-zero center-of-mass momentum and a spatially modulated order parameter arise without requiring external magnetic fields or spin imbalance. The PDW state may arise from the interplay of strong electron-electron interactions, lattice distortions, and competing orders.

PDW states are predicted to arise in a wide range of strongly correlated systems, where their formation is typically driven by the complex interplay of electron-electron interactions, lattice distortions, and competing orders such as CDWs and SDWs [6]. Despite growing theoretical interest, the microscopic origin and mechanism of PDW remain unclear from an experimental standpoint. One proposed scenario suggests that PDW order can be induced through coupling between SC states and CDW or SDW orders [6, 12], serving as a secondary order. In a conventional spin-singlet superconductor, the uniform order parameter is described by $\Delta_S(\boldsymbol{r}) = \Delta_0 e^{i\phi_S}$ ($\Delta_0$ is the amplitude of condensate wavefunction and $\phi_S$ is the phase of the order parameter). However, a unidirectional PDW introduces a spatial modulation at wavevector $\boldsymbol{P}$, modifying the order parameter to: $\Delta_{\boldsymbol{P}}(\boldsymbol{r}) =$ $\Delta(\boldsymbol{r})\, e^{i\boldsymbol{P}\cdot\boldsymbol{r}} + \Delta^*(\boldsymbol{r})\, e^{-i\boldsymbol{P}\cdot\boldsymbol{r}}$. Similarly, a unidirectional CDW modulates the charge density at wavevector $\boldsymbol{Q}$ such that $\rho_{\boldsymbol{Q}}(\boldsymbol{r}) = \rho(\boldsymbol{r})\, e^{i\boldsymbol{Q}\cdot\boldsymbol{r}} + \rho^*(\boldsymbol{r})\, e^{-i\boldsymbol{Q}\cdot\boldsymbol{r}}$. If SC states and CDW coexist, their interaction can give rise to a PDW with wavevector $\boldsymbol{Q}$ as $\Delta_{\boldsymbol{Q}}(\boldsymbol{r}) \propto \Delta_0^*(\boldsymbol{r})\rho_{\boldsymbol{Q}}(\boldsymbol{r})$ [23]. Alternatively, PDW may arise as a primary order, originating from spontaneous coherence in strongly correlated electron systems [5] or through the breaking of rotational symmetry in spin-triplet nematic phases [24]. In certain spin-liquid states or systems with pronounced spin fluctuations, Amperean pairing has also been proposed as a potential mechanism for generating a primary PDW state as proposed by Patrick A. Lee [5]. Additionally, finite momenta pairing states become the ground state of the nodal superconductor with sufficiently strong attractive nearest-neighbor interactions [25]. Interestingly, if the SC states and PDW are the predominant order in a system, two secondary CDW orders can be induced, given by : $\rho_{\boldsymbol{P}}(\boldsymbol{r}) \propto \Delta_S^*(\boldsymbol{r})\Delta_{\boldsymbol{P}}(\boldsymbol{r}) + \Delta_{-\boldsymbol{P}}^*(\boldsymbol{r})\Delta_S(\boldsymbol{r})$ and $\rho_{\boldsymbol{2P}}(\boldsymbol{r}) \propto \Delta_{-\boldsymbol{P}}^*(\boldsymbol{r})\Delta_{\boldsymbol{P}}(\boldsymbol{r})$. These systems typically exhibit complex phase diagrams where charge, magnetic, and spin orders are intricately interwoven. Although various theoretical models have been proposed to explain the emergence of PDW, a unified microscopic picture has yet to be established and remains an open question [5, 6, 9, 12, 13, 21, 24–34]. In particular, the strong coupling nature of these systems presents significant challenges, leaving the origin of PDW in many superconductors a topic of ongoing debate.

Recently, a finite-momentum Cooper pair state, termed pair density modulation (PDM), has been proposed [7, 8]. From an experimental perspective, different from the PDW that breaks long-range translational symmetry and exhibits a modulation with a long wavelength spanning several unit cells, the PDM only breaks intra-unit-cell symmetries of the space group, with its modulation period matching the lattice constant. It implies that the PDM modulates within a single unit cell, i.e. $\boldsymbol{Q}_{\text{PDM}} = \boldsymbol{Q}_{\text{Bragg}}$. However, since a pure PDW is defined as a SC state in which the order parameter exhibits spatial modulation without a uniform component [9–11], and because the symmetry of the PDW order may follow underlying crystal lattice symmetry and can be lowered by a material's microscopic detail, the PDM can, in a broad sense, be regarded as a type of PDW. The strict distinction of the point-group symmetry breaking between PDW and PDM is still under debate, requiring further theoretical and experimental study. Both PDW and PDM represent spatial inhomogeneity of the superconducting state and offer valuable insights into the complex relationship between superconductivity and competing electronic orders in strongly correlated systems.

### 1. 2 Scanning tunneling microscopy/spectroscopy

Experimentally, a variety of techniques have been employed to search for the PDW states, including angle-resolved photoemission spectroscopy, neutron scattering, X-ray scattering, nuclear magnetic resonance, and thermal transport measurements [35–41]. Among these methods, scanning tunneling microscopy and spectroscopy (STM/STS) has emerged as one of the most powerful techniques for probing the local electronic structure of materials at atomic resolution. STM is particularly effective at imaging spatial variations in the local density of states (LDOS), and offers direct insights into the spatial modulation of superconducting gap size [42]. Over the past decades, STM has played a central role in uncovering critical aspects of superconductivity, including pairing symmetry, gap inhomogeneity, dopant placement, and vortex pinning [43, 44]. The core structure for scanning in STM consists of a metallic tip approached close to a conducting sample surface, enabling the measurement of topographic image $T(\boldsymbol{r})$, tunneling current $I(\boldsymbol{r})$, and LDOS $g(\boldsymbol{r}, \mathrm{E})$.

### 1.2.1 Basic STM/STS techniques for real- and reciprocal-space characterizations

The tunneling current $I$ between the sample and tip satisfies the following relationship [42]:

$$I \propto \int_0^{eV} \rho_{\mathrm{S}}(E_{\mathrm{F}} - eV + \varepsilon)\rho_{\mathrm{T}}(E_{\mathrm{F}} + \varepsilon)\mathrm{d}\varepsilon \qquad (1)$$

where $\rho_S$ is the density of state (DOS) of the sample and $\rho_T$ is the DOS of the tip. It is important to note that the above formula is valid under the assumption that the tunneling matrix element remains relatively constant and the thermal perturbation $k_B T$ is less than the energy resolution required for the measurement.

Furthermore, if $\rho_T$ is approximately constant over the energy range of interest, then one can perform the local tunneling spectroscopy such as differential conductance ($\mathrm{d}I/\mathrm{d}V$):

$$\frac{\mathrm{d}I}{\mathrm{d}V} \propto \rho_S(E_F - eV) \qquad (2)$$

The $\mathrm{d}I/\mathrm{d}V$ spectra measured with a normal STM tip provides a direct representation of the LDOS at the sample surface.

If these $\mathrm{d}I/\mathrm{d}V$ spectra are recorded on a dense array of locations in real space, spatial variation in the LDOS is extracted and denoted as $g(\boldsymbol{r}, E = eV)$. This process, known as $\mathrm{d}I/\mathrm{d}V$ mapping, is widely employed to measure local superconducting energy gap variations and to visualize vortex structures under a magnetic field [45]. Besides, 'Setup effect', a systematic error that possibly results in a misidentification of the energy range where the density of states shows spatial modulations, can be suppressed by using the following formula:

$$Z(\boldsymbol{r}, E) = g(\boldsymbol{r}, E)/g(\boldsymbol{r}, -E) \qquad (3)$$

Beyond mapping spatial variation of LDOS, STM/STS also enables quantitative characterization of the energy-resolved electronic properties in superconductors. The LDOS of superconducting state is characterized by the two pronounced coherence peaks positioned near the Fermi energy, which are associated with the formation of the SC energy gap. This energy gap, typically denoted as Δ, is related to the superconducting order parameter. STM/STS serves as a powerful and reliable tool for estimating the SC gap and provides critical insights into the underlying electronic properties of the superconducting material [44].

In addition, the Fourier transform (FT) spectrum, which can transform the physical quantities from real space into momentum space, is widely utilized in experimental data analysis [46, 47]. By applying this technique in STM/STS to transform the differential conductance map from real space to $\boldsymbol{q}$-space, electronic structures can be probed in detail, including dominant scattering vectors which are often associated with physical phenomena in the sample, such as quasiparticle interference (QPI) and electronic state modulations. Importantly, these local features are challenging to access using other bulk measurement techniques, such as neutron scattering or thermal transport measurements [8]. These foundational STM/STS techniques provide the necessary framework for identifying experimental signatures of PDW states.

### 1.2.2 Several signatures of PDW from STM/STS measurements

By measuring the differential conductance variations in real space, STM can reveal spatially varying electronic patterns that may correspond to the modulations expected from PDW order. This makes STM suited for the detection of PDW, offering direct insight into the real-space manifestations of superconducting states. Specifically, STM can measure the following physical quantities to characterize the spatial modulation.

(i) Spatial modulation of the SC energy gap $\Delta(\boldsymbol{r})$

A "gap map" $\Delta(\boldsymbol{r})$ can be obtained by extracting superconducting energy gap $\Delta$ for every spectrum within a DOS map, providing direct visualization of spatial variation of SC energy gap as $\delta\Delta(\boldsymbol{r}) = \Delta(\boldsymbol{r}) - \Delta_0(\boldsymbol{r})$, where $\Delta_0(\boldsymbol{r})$ is the spatially averaged value.

Considering that the magnitude of the SC energy gap $\Delta$ in the sample is determined by the energy scale of the pairing interactions, the spatial periodic modulation of the superconducting energy gap provides signatures of the presence of a PDW, as described by the following formula [2, 3]:

$$\Delta(\boldsymbol{r}) = \Delta_\mathrm{S} + \Delta_\mathrm{P}\cos(\boldsymbol{P}\cdot\boldsymbol{r}) \quad (4)$$

where $\Delta_\mathrm{S}$ is the homogeneous SC energy gap, and $\Delta_\mathrm{P}\cos(\boldsymbol{P}\cdot\boldsymbol{r})$ represents the PDW component with wavevector $\boldsymbol{P}$.

Determining the energy gap in materials precisely is challenging because thermal and phase fluctuations may destroy coherence, resulting in broadened and indeterminate coherence peaks [23]. Superconducting tips with sharp coherence peaks and certain energy gap $\Delta_\mathrm{tip}$ are employed to overcome this difficulty. Through convolution between the DOS of the tip and the sample, the resolution for detecting the sample's energy gap, marked by the maxima in $N_\mathrm{s}(E)$, is significantly enhanced. This process shifts the energy positions of these maxima to $E = \pm(\Delta_\mathrm{s}+\Delta_\mathrm{tip})$, enabling accurate extraction of the sample's energy gap. Moreover, the extrema of the negative second derivative of d*I*/d*V*, defined as $D(\boldsymbol{r},E) \equiv -\mathrm{d}^3I/\mathrm{d}V^3(\boldsymbol{r},E)$, can determine the coherence peak positions precisely, thereby providing a robust measurement of the energy gap size. This quantity has also been shown to be sensitive to the strength of superconductivity, as reported by Ruan *et al.*[48].

(ii) Spatial modulation of condensed electron-pair density $\boldsymbol{n}(r)$

Besides precise measurement of energy gap, superconducting tips can also be applied to form a superconductor-insulator-superconductor (SIS) junction, where the condensed electron-pair density at location $\boldsymbol{r}$, denoted as $n(\boldsymbol{r})$, can be visualized by measuring the tip-sample Josephson critical-current squared $I_J^2(\boldsymbol{r})$ and the normal-state junction resistance $R_{\mathrm{N}}$. These quantities are related via:

$$n(\boldsymbol{r}) \propto I_{\mathrm{J}}^2(\boldsymbol{r})R_{\mathrm{N}}^2(\boldsymbol{r}) \tag{5}$$

The spatial modulation of condensed electron-pair density $n(\boldsymbol{r})$ serves as another key signature of PDW states. This technique is known as scanned Josephson tunnelling microscopy (SJTM), and in the experimental scenario where the Josephson coupling energy $E_J$ is less than the thermal fluctuation energy $k_BT$, i.e. $E_J < k_BT$, the Josephson junction enters a phase-diffusive steady state when a voltage $V$ is applied. In this condition, the electron-pair current $I_{\mathrm{CP}}(V)$ adheres to the following expression [49–51]:

$$I_{\mathrm{CP}}(V) = \frac{I_{\mathrm{J}}^2 ZV}{2(V^2+V_c^2)} \tag{6}$$

where $V_{\mathrm{c}}$ represents the thermal fluctuations as Johnson noise generated by a resistor $Z$. From Formula (6), the following expression can be derived:

$$\mathrm{d}I_{\mathrm{CP}}/\mathrm{d}V \equiv g(V) = \frac{I_{\mathrm{J}}^2 Z(V_c^2-V^2)}{2(V^2+V_c^2)^2} \tag{7}$$

From Formula (7), it follows that $g(0) \approx I_{\mathrm{J}}^2$, and the maximum value of the electron-pair current $I_m \propto I_{\mathrm{J}}^2$. Therefore, the electron-pair density can be expressed as:

$$N_{CP} = n(r) \approx g(0)R_{\mathrm{N}}^2(\boldsymbol{r}) \tag{8}$$

It is important to note that the presence of a homogeneous condensate leads to a modulated Josephson critical current:

$$I_{\mathrm{J}}(\boldsymbol{r}) = I_{J_s} + I_{J_p}\cos(\boldsymbol{P}\cdot\boldsymbol{r}) \tag{9}$$

where the first term $I_{J_s}$ represents Cooper-pair tunnelling to a homogeneous condensate and the second $I_{J_p}\cos(\boldsymbol{P}\cdot\boldsymbol{r})$ represents contributions from the PDW with wavevector $\boldsymbol{P}$. The two components are independent when considering momentum conservation and the modulation period of the maximum value of electron-pair current $I_m$ differs from that of the PDW order [49, 52–54].

Notably, if the two components are coherent, the modulation period of $I_m$ is consistent with $\Delta(\boldsymbol{r})$.

(iii) Spatial modulation of coherence peak height

The coherence peak height is positively correlated with the superconducting order parameter [48]: a higher coherence peak in STS usually reflects a higher superfluid density. Moreover, the sharpness of coherence peaks, which can be characterized by the peak in $D(\boldsymbol{r}, E)$, indicates the strength of superconducting coherence [7]. However, it warrants special attention that the relationship between coherence peak heights and superfluid density remains unclear; conclusions regarding coherence peaks primarily stem from empirical observations [48, 55–59].

(iv) Phase-winding of CDW order

If PDW is a primary order with wavevector $\boldsymbol{P}$ which induces a CDW order with wavevector $2\boldsymbol{P}$, then a π-phase shift in the PDW order around a half-dislocation is predicted to nucleate at a topological defect with a $2\pi$ phase-winding in the induced $2\boldsymbol{P}$ CDW order [54, 60–62]. Such a topological relation between PDW and CDW orders can be explored through STM imaging of phase textures. This may be facilitated by the numerical 2D lock-in technique, which allows spatial mapping of the local phase of density wave orders and detection of associated topological defects.

Exploration of the magnitude and phase of the modulation can be achieved through the 2D lock-in technique [54]. Let $A(\boldsymbol{r}) = \sum_{\boldsymbol{Q}} a_{\boldsymbol{Q}}(\boldsymbol{r})\, e^{i\boldsymbol{Q}\cdot\boldsymbol{r}}$ denote an arbitrary image in real space, with $\boldsymbol{Q}$ representing the target wavevector and $a_{\boldsymbol{Q}}(\boldsymbol{r})$ representing complex amplitude at wavevector $\boldsymbol{Q}$ and position $\boldsymbol{r}$. It follows that:

$$A_{\boldsymbol{Q}}(\boldsymbol{r}) = \int A(\boldsymbol{R}) e^{i\boldsymbol{Q}\cdot\boldsymbol{R}} e^{-\frac{(\boldsymbol{r}-\boldsymbol{R})^2}{2\sigma_r^2}} \mathrm{d}\boldsymbol{R} \tag{10}$$

$$A_{\boldsymbol{Q}}(\boldsymbol{r}) = \mathcal{F}^{-1}\left(A_{\boldsymbol{Q}}(\boldsymbol{q})\right) = \mathcal{F}^{-1}\left[\mathcal{F}\left(A(\boldsymbol{r}) e^{i\boldsymbol{Q}\cdot\boldsymbol{r}}\right) \cdot \frac{1}{\sqrt{2\pi}\sigma_q} e^{-\frac{\boldsymbol{q}^2}{2\sigma_q^2}}\right] \tag{11}$$

$$\left|A_{\boldsymbol{Q}}(\boldsymbol{r})\right| = \sqrt{[\mathrm{Re}A_{\boldsymbol{Q}}(\boldsymbol{r})]^2 + [\mathrm{Im}A_{\boldsymbol{Q}}(\boldsymbol{r})]^2} \tag{12}$$

$$\Phi_{\boldsymbol{Q}}^{A}(\boldsymbol{r}) = \tan^{-1}\left(\frac{\mathrm{Im}A_{\boldsymbol{Q}}(\boldsymbol{r})}{\mathrm{Re}A_{\boldsymbol{Q}}(\boldsymbol{r})}\right) \tag{13}$$

where $\mathcal{F}$ denotes Fourier transform, $\sigma_r$ denotes the cut-off length in $\boldsymbol{r}$-space and $\sigma_q$ corresponds to the cut-off length in $\boldsymbol{q}$-space. Formula (10, 11) represent approximate complex amplitude calculated in real space and $\boldsymbol{q}$-space, and Formula (12, 13) represent the spatial magnitude and phase of $A_{\boldsymbol{Q}}(\boldsymbol{r})$. In practice, Formula (10, 11) can be used to filter the map with a modulation wavevector $\boldsymbol{Q}$, yielding a real space map that retains only the modulation components near $\boldsymbol{Q}$. This procedure enables a direct visualization of the spatial modulation associated with wavevector $\boldsymbol{Q}$. For instance, by analyzing the spatial variation amplitude and phase of energy gap map $\Delta(\boldsymbol{r})$ at wavevector $\boldsymbol{Q}$, one can obtain the filtered energy gap map $\Delta_{\boldsymbol{Q}}(\boldsymbol{r})$, spatial magnitude $\left|A_{\boldsymbol{Q}}^{\Delta}(\boldsymbol{r})\right|$ and phase $\Phi_{\boldsymbol{Q}}^{\Delta}(\boldsymbol{r})$ using the 2D lock-in technique. Moreover, the nematicity of the density wave state can be determined through calculating the nematic order parameter $F(\boldsymbol{r})$:

$$F(\boldsymbol{r}) = \frac{|A_{\boldsymbol{Q}_x}(\boldsymbol{r})| - |A_{\boldsymbol{Q}_y}(\boldsymbol{r})|}{|A_{\boldsymbol{Q}_x}(\boldsymbol{r})| + |A_{\boldsymbol{Q}_y}(\boldsymbol{r})|} \tag{14}$$

However, the 2D lock-in technique is a numerical post-processing method rather than a direct experimental observation. Its results depend sensitively on parameters such as the cut-off length σ, filtering window, and target wavevector $\boldsymbol{Q}$. Therefore, while it offers valuable supporting insights, it must be interpreted with caution.

In addition, in some unconventional superconductors, especially cuprates where $CuO_2$ planes play a crucial role in enabling superconductivity, *d*-symmetry sublattice-phase-resolved Fourier analysis is essential to determine the symmetry of form factors. Taking cuprates as an example, one can calculate the sublattice-phase-resolved $Z(\boldsymbol{r}, E)$ image through Formula (3) and decompose it into three components: $Cu(\boldsymbol{r})$ representing the measured $Z(\boldsymbol{r})$ values at Cu sites, and $O_x(\boldsymbol{r})$ and $O_y(\boldsymbol{r})$ corresponding to the measurements at oxygen sites along the *x*- and *y*- axes, respectively. The phase-resolved Fourier transforms of $O_x(\boldsymbol{r})$ and $O_y(\boldsymbol{r})$, denoted as $\tilde{O}_x(\boldsymbol{q})$ and $\tilde{O}_y(\boldsymbol{q})$, are then employed to analyze the form factor symmetry for modulations at any given wavevector $\boldsymbol{q}$. For a density wave at wavevector $\boldsymbol{Q}$, the magnitudes of the associated form factors can be evaluated [63], including the *d*-symmetry form factor $\tilde{D}^Z(\boldsymbol{q})$ :

$$\tilde{D}^Z(\boldsymbol{q}) = \tilde{O}_x(\boldsymbol{q}) - \tilde{O}_y(\boldsymbol{q}) \tag{15}$$

*s'*-symmetry form factor $\tilde{S}'^Z(\boldsymbol{q})$ :

$$\tilde{S}'^Z(\boldsymbol{q}) = \tilde{O}_x(\boldsymbol{q}) + \tilde{O}_y(\boldsymbol{q}) \tag{16}$$

and *s*-symmetry form factor $\tilde{S}^Z(\boldsymbol{q})$ :

$$\tilde{S}^Z(\boldsymbol{q}) = \widetilde{Cu}(\boldsymbol{q}) \tag{17}$$

where the superscript $Z$ identifies the type of sublattice-resolved data used. Then, modulations at any $\boldsymbol{Q}$ having *d/s/s'*- symmetry form factor generate a peak in $\tilde{D}^Z(\boldsymbol{q})/\tilde{S}^Z(\boldsymbol{q})/\tilde{S}'^Z(\boldsymbol{q})$ at wavevector $\boldsymbol{Q}$ [64].

(v) Spatial modulation of Andreev reflection

When superconducting tips are used, Andreev reflection will exhibit modulation as well [23]. If a significant difference exists between the superconducting energy gaps of the tip and the sample, the Andreev reflection process will introduce more complex tunneling dynamics. This is particularly evident when the edge of the smaller energy gap aligns with the chemical potential of the other superconductor, enhancing the Andreev reflection and resulting in the emergence of pronounced energy peaks in the d*I*/d*V* spectrum [65]. Furthermore, measurements of Andreev resonance reveal significant spatial modulation, highlighting implications of PDW states [23].

Taken together, these STM-based signatures constitute an experimental framework for identifying PDW states and probing their intricate interplay with coexisting or competing electronic orders. Building upon this methodological foundation, the following section provides a comprehensive review of experimentally observed PDW phenomena across diverse families of superconductors.

## 2. Experimental observations of PDW/M in Superconductors

Since PDW was first proposed in 2002 [30], extensive theoretical and experimental efforts have been devoted to exploring this phenomenon. Recent studies using STM have revealed distinct signatures of PDW states, their interplay with competing orders, and their microscopic origins in various material systems, including cuprate superconductors, iron-based superconductors, spin-triplet superconductors, Kagome lattice superconductors, and transition metal dichalcogenides (TMDs), as detailed in Table 1. Considering the extensive scope of PDW research across these systems, this review endeavors to highlight key findings and summarize STM's contributions to studies on PDW.

Table 1: Superconductors Exhibiting PDW/PDM Observed via STM.

| Types | Materials | $T_c$ | Wavevectors of PDW/PDM states | Wavevectors of CDW states | Technique | Ref. |
|---|---|---|---|---|---|---|
| Cuprate superconductors | $Bi_2Sr_2CaCu_2O_{8+x}$ ($p$ = 17%, crystals) | 88 K | $\boldsymbol{Q}_{2q-4a_0} = \frac{1}{4}\boldsymbol{Q}^{Cu}_{Bragg}$ | $\boldsymbol{Q}_{1q-a_0/0.22} = 0.22\ \boldsymbol{Q}^{Cu}_{Bragg}$ | SJTM | [52] |
| | $Bi_2Sr_2CaCu_2O_{8+x}$ ($p \approx$ 17%, crystals) | 88 K | $\boldsymbol{Q}_{2q-8a_0} = \frac{1}{8}\boldsymbol{Q}^{Cu}_{Bragg}$ | $\boldsymbol{Q}_{2q-8a_0} = \frac{1}{8}\boldsymbol{Q}^{Cu}_{Bragg}$ <br> $\boldsymbol{Q}_{2q-4a_0} = \frac{1}{4}\boldsymbol{Q}^{Cu}_{Bragg}$ | STM | [64] |
| | $Bi_2Sr_2CaCu_2O_{8+x}$ ($p \approx$ 6%, crystals) | 10 K | $\boldsymbol{Q}_{1q-a_0/0.28} = 0.28\ \boldsymbol{Q}^{Cu}_{Bragg}$ | $\boldsymbol{Q}_{1q-a_0/0.27} = 0.27\ \boldsymbol{Q}^{Cu}_{Bragg}$ | STM | [48] |
| | $Bi_2Sr_2CaCu_2O_{8+x}$ ($p \approx$ 17%, crystals) | 91 K | $\boldsymbol{Q}_{2q-8a_0} = \frac{1}{8}\boldsymbol{Q}^{Cu}_{Bragg}$ | $\boldsymbol{Q}_{2q-8a_0} = \frac{1}{8}\boldsymbol{Q}^{Cu}_{Bragg}$ <br> $\boldsymbol{Q}_{2q-4a_0} = \frac{1}{4}\boldsymbol{Q}^{Cu}_{Bragg}$ | SJTM | [54] |
| | $Bi_2Sr_2CaDyCu_2O_8$ | 37±3 K | $\boldsymbol{Q}_{2q-8a_0} = \frac{1}{8}\boldsymbol{Q}^{Cu}_{Bragg}$ | $\boldsymbol{Q}_{2q-8a_0} = \frac{1}{8}\boldsymbol{Q}^{Cu}_{Bragg}$ | STM | [66] |
| | ($p \approx 8\%$, crystals) | | | $\boldsymbol{Q}_{2q-4a_0} = \frac{1}{4}\boldsymbol{Q}_{\mathrm{Bragg}}^{\mathrm{Cu}}$ | | |
| | $Bi_2Sr_2CaCu_2O_{8+x}$ ($p \approx 17\%$, crystals) | ------ | $\boldsymbol{Q}_{2q-4a_0} = \frac{1}{4}\boldsymbol{Q}_{\mathrm{Bragg}}^{\mathrm{Cu}}$ | ------ | SJTM | [53] |
| | $Bi_{2.08}Sr_{1.92}CuO_{6+\delta}$ (overdoped, crystals) | 7 K | $\boldsymbol{Q}_{2q-4a_0/3} = \frac{3}{4}\boldsymbol{Q}_{\mathrm{Bragg}}^{\mathrm{Cu}}$ | ------ | STM | [67] |
| Iron-based superconductors | $FeTe_{1-x}Se_x$/STO (x ≈ 0.5, films) | ------ | $\boldsymbol{Q}_{1q-a_{\mathrm{Fe}}/0.28} = 0.28\boldsymbol{Q}_{\mathrm{Bragg}}^{\mathrm{Fe}}$ | $2\boldsymbol{Q}_{1q-a_{\mathrm{Fe}}/0.28} = 0.56\boldsymbol{Q}_{\mathrm{Bragg}}^{\mathrm{Fe}}$ | STM | [61] |
| | $FeTe_{1-x}Se_x$/STO (x ≈ 0.7, films) | ------ | $\boldsymbol{Q}_{2q-4a_{\mathrm{Fe}}} = \frac{1}{4}\boldsymbol{Q}_{\mathrm{Bragg}}^{\mathrm{Fe}}$ | $\boldsymbol{Q}_{2q-4a_{\mathrm{Fe}}} = \frac{1}{4}\boldsymbol{Q}_{\mathrm{Bragg}}^{\mathrm{Fe}}$ | STM | [68] |
| | $EuRbFe_4As_4$ (crystals) | 37 K | $\boldsymbol{Q}_{1q-8a_{\mathrm{Fe}}} = \frac{1}{8}\boldsymbol{Q}_{\mathrm{Bragg}}^{\mathrm{Fe}}$ | ------ | STM | [69] |
| | $FeTe_{1-x}Se_x$/STO (x ≈ 0.7, films) | 58 K | $\boldsymbol{Q}_{2q-a_{\mathrm{Te,Se}}} = \boldsymbol{Q}_{\mathrm{Bragg}}^{\mathrm{Te,Se}}$ | ------ | STM | [7] |
| | $FeTe_{1-x}Se_x$ (x = 0.45, thin flakes) | 14.5 K | $\boldsymbol{Q}_{2q-a_{\mathrm{Fe}}} = \boldsymbol{Q}_{\mathrm{Bragg}}^{\mathrm{Fe}}$ | ------ | STM | [8] |
| | FeSe/STO (films) | ------ | $\boldsymbol{Q}_{1q-5.4a_{\mathrm{Fe}}} = \frac{1}{5.4}\boldsymbol{Q}_{\mathrm{Bragg}}^{\mathrm{Fe}}$; $\boldsymbol{Q}_{2q-\sqrt{2}a_{\mathrm{Fe}}} = \frac{1}{\sqrt{2}}\boldsymbol{Q}_{\mathrm{Bragg}}^{\mathrm{Fe}}$ | ------ | SJTM | [70] |
| Spin-triplet superconductors | $UTe_2$ (crystals) | 1.65 K | Incommensurate $\boldsymbol{P}_{1,2,3}$ | Almost same as PDW, $\boldsymbol{Q}_{1,2,3}$ | SJTM | [23] |
| Kagome lattice superconductors | $CsV_3Sb_5$ (crystals) | 2.3 K | $\boldsymbol{Q}_{3q-4a/3} = \frac{3}{4}\boldsymbol{Q}_{\mathrm{Bragg}}$ | $\boldsymbol{Q}_{1q-4a} = \frac{1}{4}\boldsymbol{Q}_{\mathrm{Bragg}}$; $\boldsymbol{Q}_{3q-2a} = \frac{1}{2}\boldsymbol{Q}_{\mathrm{Bragg}}$ | STM | [71] |
| | $KV_3Sb_5$ (crystals) | 0.93 K | $\boldsymbol{Q}_{3q-2a} = \frac{1}{2}\boldsymbol{Q}_{\mathrm{Bragg}}$ | $\boldsymbol{Q}_{3q-2a} = \frac{1}{2}\boldsymbol{Q}_{\mathrm{Bragg}}$ | SJTM and STM | [72] |
| | $CsV_3Sb_5$ | 2.3K (bulk) | $\boldsymbol{Q}_{3q-4a} = \frac{1}{4}\boldsymbol{Q}_{\mathrm{Bragg}}$ | $\boldsymbol{Q}_{3q-2a} = \frac{1}{2}\boldsymbol{Q}_{\mathrm{Bragg}}$ | STM | [73] |
| | (crystals) | 5.4 K(quasi-2D) | | | | |
| TMDs superconductors | $NbSe_2$ (crystals) | ------ | $\boldsymbol{Q}_{3\text{q}-3a} = \frac{1}{3}\boldsymbol{Q}_{\text{Bragg}}$ | $\boldsymbol{Q}_{3\text{q}-3a} \approx \frac{1}{3}\boldsymbol{Q}_{\text{Bragg}}$ (incommensurate) | SJTM | [74] |
| | $NbSe_2$ (crystals) | 7.2 K | $\boldsymbol{Q}_{3\text{q}-3a} = \frac{1}{3}\boldsymbol{Q}_{\text{Bragg}}$ | $\boldsymbol{Q}_{3\text{q}-3a} \approx \frac{1}{3}\boldsymbol{Q}_{\text{Bragg}}$ (incommensurate) | STM | [75] |
| | $MoTe_2$ (films) | 6.0 K | $\boldsymbol{Q}_{1\text{q}-5a} = \frac{1}{5}\boldsymbol{Q}_{\text{Bragg}}$ | $\boldsymbol{Q}_{1\text{q}-5a} = \frac{1}{5}\boldsymbol{Q}_{\text{Bragg}}$; $\boldsymbol{Q}_{1\text{q}-5a/2} = \frac{2}{5}\boldsymbol{Q}_{\text{Bragg}}$ | STM | [76] |
| | $MoTe_2$ (films) | 6.0 K | $\boldsymbol{Q}_{1\text{q}-a} = \boldsymbol{Q}^{a}_{\text{Bragg}}$; $\boldsymbol{Q}_{1\text{q}-a/2} = 2\boldsymbol{Q}^{a}_{\text{Bragg}}$; $\boldsymbol{Q}_{1\text{q}-3b} = \frac{1}{3}\boldsymbol{Q}^{b}_{\text{Bragg}}$ | $\boldsymbol{Q}_{1\text{q}-2a} = \frac{1}{2}\boldsymbol{Q}^{a}_{\text{Bragg}}$; $\boldsymbol{Q}_{1\text{q}-5b} = \frac{1}{5}\boldsymbol{Q}^{b}_{\text{Bragg}}$; $\boldsymbol{Q}_{1\text{q}-3b} = \frac{1}{3}\boldsymbol{Q}^{b}_{\text{Bragg}}$; $\boldsymbol{Q}_{1\text{q}-a/2} = 2\boldsymbol{Q}^{a}_{\text{Bragg}}$ | STM | [77] |

In Table 1, $a_0$, $a_{\text{Fe}}$, $a_{\text{Te,Se}}$, $a$ and $b$ denote the lattice constants, while $p$ represents the hole density in the cuprates. The subscripts in the wavevectors $\boldsymbol{Q}_{n\text{q}-ma}$ encode the characteristics of the PDW states: specifically, '$n$' indicates the number of directions in which spatial modulations occur, '$m$' denotes the modulation periodicity, and '$a$' or '$b$' refers to the corresponding lattice constant. The term $\boldsymbol{Q}^{(a,b)}_{\text{Bragg}}$ represents the Bragg vectors of the sample, with the superscript indicating the specific crystallographic direction (if the superscript represents an element $A$, it indicates that the corresponding Bragg wavevector is the Bragg peaks of $A$ lattice within the crystal). For example, in Kagome lattice superconductors $CsV_3Sb_5$, the wavevectors $\boldsymbol{Q}_{3\text{q}-4a/3} = \frac{3}{4}\boldsymbol{Q}^{(a,b)}_{\text{Bragg}}$ describes a bidirectional 3Q electronic modulation with a periodicity of $4a/3$ along Bragg vectors, which remains consistent across two different crystallographic axes.

## 2.1 Cuprate Superconductors

Since their discovery in 1986, cuprate superconductors have become a central topic in condensed matter physics due to their remarkably high superconducting transition temperatures $T_c$ and complex phase diagrams [78–82]. Structurally, cuprates consist of layered crystal lattices where $CuO_2$ planes host strongly correlated electrons, resulting in the emergence of various phenomena such as CDWs [83], SDWs [84, 85], pseudogap [79, 80], and PDWs [33, 86, 87]. Recent studies have proposed that the 'pseudogap' regime contains an unconventional density wave with a $d$- symmetry form factor in high-temperature superconducting cuprate [63, 88], suggesting the presence of a PDW intertwined with other electronic orders. The emergence of PDW states in cuprates not only reflects the complexity of their correlated electron landscape but also offers a potential key to understanding the pseudogap phase and superconducting mechanism in cuprates.

Numerous theoretical studies have supported the presence of PDW in cuprates [30–33, 86, 89, 90], and the existence of PDW can explain various phenomena, such as the absent *c*-axis superconductivity in $La_{2-x}Ba_xCuO_4$ [87], unusual characteristics in the single-particle excitations, and the cuprate pseudogap phase [5, 81, 91].

Hamidian *et al.* were the first to successfully visualize Cooper-pair density modulations in near-optimally doped $Bi_2Sr_2CaCu_2O_{8+x}$ (Bi-2212, $p$ = 17%) using SJTM [52]. They achieved this by attaching a nanometer-scale Bi-2212 flake to a tungsten tip, effectively creating a *d*-wave superconducting tip. The observed modulations in $I_m(\boldsymbol{r})$ (Formula 7) at wavevector $\boldsymbol{Q}_{2\text{q}-4a_0} = \frac{1}{4}\boldsymbol{Q}_{\text{Bragg}}^{\text{Cu}}$ provided signatures consistent with a PDW coexisting with a robust Cooper-pair condensate in Bi-2212, as shown in Figure 2 (a,b). Analysis of the oxygen sublattice-phase-resolved *d*-symmetry form factor density modulations [88] reveals that the CDW displays a *d*-symmetry form factor, while the PDW exhibits an *s/s'*-symmetry form factor [63]. This suggests that the PDW with wavevector $\boldsymbol{Q}_{2\text{q}-4a_0}$ may arise from the coupling between superconductivity and the CDW, whose wavevector $\boldsymbol{Q}_{\text{C}}$ is approximately equal to $\boldsymbol{Q}_{2\text{q}-4a_0}$.

Novel electronic phases may appear under high magnetic fields which suppress the superconductivity in underdoped cuprates, and many studies tend to link the electronic phases to PDW [5]. In particular, Edkins *et al.* reported the signatures of magnetic-field-induced PDW states in slightly underdoped $Bi_2Sr_2CaCu_2O_{8+x}$ [64]. Measurements of field-induced effects on electronic structure revealed modulations of differential conductance with wavevectors $\boldsymbol{Q}_{2\text{q}-8a_0} = \frac{1}{8}\boldsymbol{Q}_{\text{Bragg}}^{\text{Cu}}$ and $2\boldsymbol{Q}_{2\text{q}-8a_0}$, which were confined to the vortex halo regions (Figure 2 (c,d)). The SC energy gap and sublattice-phase-resolved Fourier analysis suggested the presence of a magnetic-field-induced bidirectional PDW with wavevector $\boldsymbol{Q}_{2\text{q}-8a_0}$ and *d*-symmetry form factor. The authors proposed that this PDW constitutes a primary order, which coexists with secondary CDW modulations at $\boldsymbol{Q}_{2\text{q}-8a_0}$ and $2\boldsymbol{Q}_{2\text{q}-8a_0}$, aligning with theoretical predictions. Based on these observations, they further suggested that the high-field phase of cuprates may be dominated by PDW states, with an intertwined CDW component [5], thereby offering critical insights into the interplay between superconductivity and competing ordered phases in cuprates.

The temperature-hole doping phase diagram in cuprates highlights the critical role of doping in shaping their electronic properties [92]. Ruan et al. extended their research to severely underdoped Bi-2212, identifying PDWs using a novel method based on the analysis of superconducting coherence peaks and gap depth [48]. The tunneling current $I(\boldsymbol{r})$ map exhibited a distinct chequerboard pattern with a modulation wavevector $\boldsymbol{Q}_{1\text{q}-a_0/0.27} = 0.27\,\boldsymbol{Q}_{\text{Bragg}}^{\text{Cu}}$ along the Cu–Cu bond. The negative second derivative of d$I$/d$V$, $D(\boldsymbol{r})$, and coherence peak height $H(\boldsymbol{r})$ further revealed clear spatial modulations in superconducting coherence with wavevector $\boldsymbol{Q}_{1\mathrm{q}-a_0/0.28} = 0.28\ \boldsymbol{Q}_{\mathrm{Bragg}}^{\mathrm{Cu}}$, as shown in Figure 2 (e-g). These observations, along with the cross-correlation analysis, demonstrate a positive correlation between charge order and pair-density order. These findings align with models predicting the coexistence of intertwined PDW, $d$-form factor charge order, and $d$-wave superconductivity.

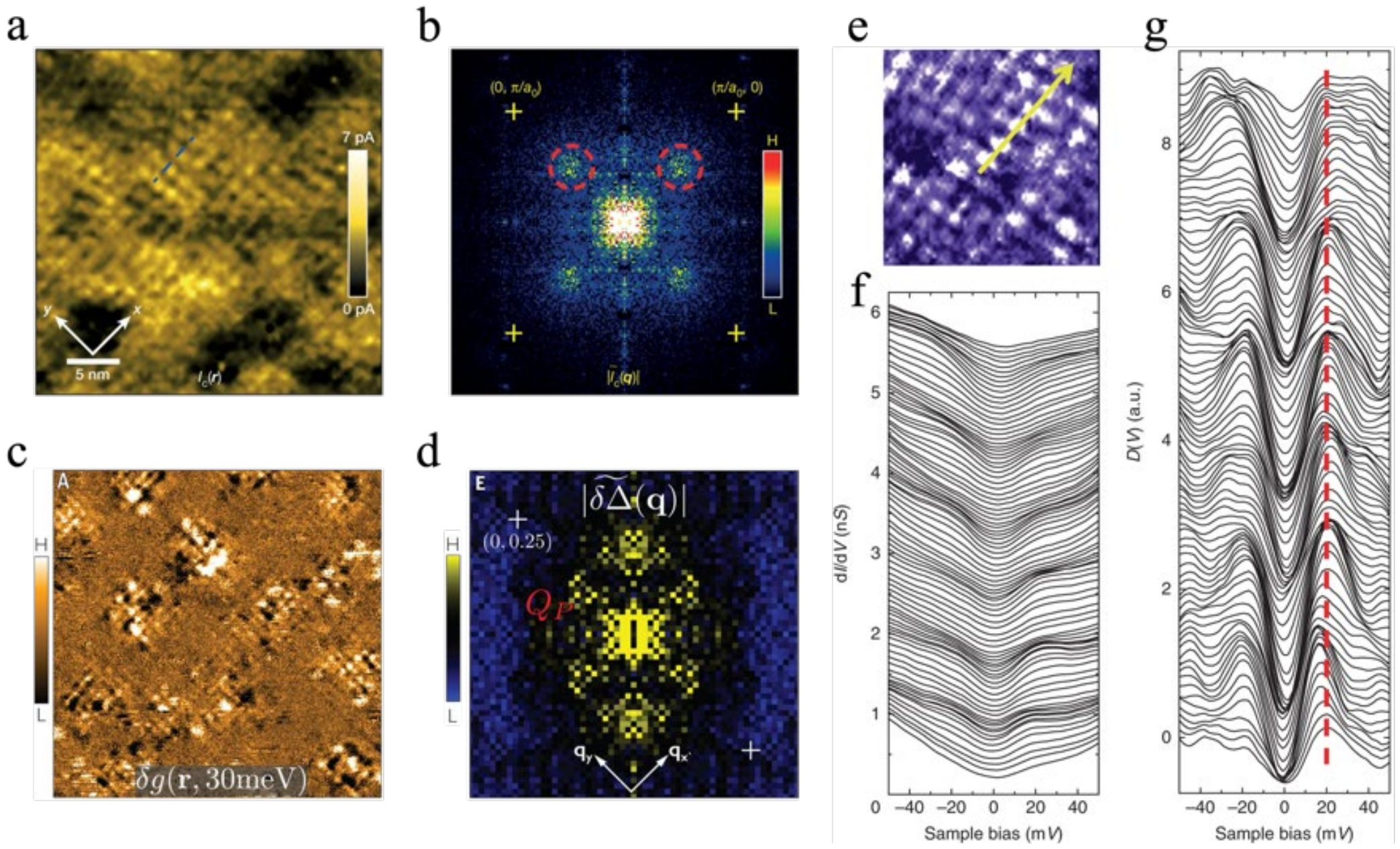

**Fig. 2** Initial experimental indications of PDW in Bi-2212. (**a**-**b**) Modulation of $I_m(\boldsymbol{r})$ in near-optimally doped Bi-2212 [52]. (a) Maximum value of electron-pair current map $I_m(\boldsymbol{r})$. (The notation of maximum value of electron-pair current used in this figure is $I_c$.) (b) The FT of (a) shows clear modulation wavevectors at $\boldsymbol{Q}_{2\mathrm{q}-4a_0}$ (marked by red dashed circles). (**c**-**d**) Field-induced PDW in slightly underdoped Bi-2212 [64]. (c) The field-induced modulations δg($\boldsymbol{r}$, 30 m$e$V) = g($\boldsymbol{r}$, 30 m$e$V, B = 8.25 T) − g($\boldsymbol{r}$, 30 m$e$V, B = 0 T). (d) FT of SC energy gap variations $\delta\Delta(\boldsymbol{r}) = \Delta(\boldsymbol{r}$, 8.25 T) − $\Delta(\boldsymbol{r}$, 0 T). The wavevector of field induced gap modulation is $\boldsymbol{Q}_{\mathbf{P}} = \boldsymbol{Q}_{2\mathrm{q}-8a_0}$. (**e**-**g**) Spatial modulation of coherence peak in severely underdoped Bi-2212 [48]. (e) The current map $I(\boldsymbol{r})$. (f) The linecut along the trajectory indicated as the yellow arrow in panel (e). (g) $D(\boldsymbol{r})$, numerical calculation of the negative second derivative of scanning tunnelling spectroscopy curves in panel (f).

However, the aforementioned works primarily focused on the analysis of tunneling current, superconducting coherence peaks, and energy gap depth, or offered a preliminary discussion on variations in the SC energy gap $\Delta(\boldsymbol{r})$, lacking a comprehensive examination of the periodic modulation of $\Delta(\boldsymbol{r})$. Moreover, with the advancement of numerical 2D lock-in technique, phase-resolved analysis has become feasible, offering new insights into the spatial structure and topological properties of PDW states. Du *et al.* investigated a PDW that induced a secondary CDW coexisting with superconductivity in nearly optimally doped Bi-2212 by directly analyzing the energy gap [54]. Spatially modulated $\Delta(\boldsymbol{r})$ at wavevector $\boldsymbol{Q}_{2\mathrm{q}-8a_0} = \frac{1}{8}\boldsymbol{Q}_{\mathrm{Bragg}}^{\mathrm{Cu}}$, combined with 2D lock-in analysis, revealed that the PDW state in cuprates was microscopically unidirectional and potentially existed in a vestigial nematic state (Figure 3 (a-c)). The phase-resolved imaging revealed that topological defects with $2\pi$ phase winding in the induced $2\boldsymbol{Q}_{2\mathrm{q}-8a_0}$ density wave coincided spatially with $\pi$ phase shifts in the PDW order (Figure 3 (d)), indicating a robust interplay between the CDW topological defects and the local PDW phase structure, consistent with theoretical expectations for the interaction between induced CDW dislocations and PDW order [60].

At certain doping levels, copper oxides undergo a transition from the SC states into the pseudogap phase with increasing temperature [92]. Accordingly, temperature-dependent experiments provide an effective means to research the relationship between the PDW state and the pseudogap in cuprates. Wang *et al.* provided a detailed visualization of the temperature-dependent evolution of electronic structures in severely underdoped $Bi_2Sr_2CaDyCu_2O_8$ [66]. This material exhibited $8a_0$ periodic $\Delta(\boldsymbol{r})$ modulations, indicative of a PDW coexisting with superconductivity. As the temperature rose and the system entered the pseudogap phase, these modulations persisted but became thermally broadened, undergoing minimal changes, as shown in Figure 3 (e,f). Furthermore, the QPI pattern transitioned from matching the predicted $8a_0$-periodic PDW in the superconducting state to that of a pure $8a_0$-periodic PDW in the pseudogap phase, beyond $T_c$, marking a distinct difference from the $4a_0$-periodic CDW. This suggests that the pseudogap phase of $Bi_2Sr_2CaDyCu_2O_8$ harbors a PDW state, which is predominantly influenced by quantum phase fluctuations.

Many theoretical studies have predicted that a striped PDW state can spontaneously emerge in cuprates, driven by strong electronic correlations, rather than being induced by CDW [30, 93]. Building on this theoretical framework, Chen *et al.* reported signatures of a nematic PDW state [53] in Bi-2212 ($p \approx 17\%$) in the absence of detectable CDW order. The observed PDW state was interpreted as a primary electronic order, exhibiting locally unidirectional and $4a_0$ lattice-commensurate electron-pair density modulations. This represents a fundamental form of symmetry breaking in hole-doped $CuO_2$ planes, where electron-pair density modulations spontaneously generate a vestigial nematic phase, independent of CDW order. Inspired by the previous study [94] on $La_{2-x}Ba_xCuO_4$, they imaged the Zn impurities and inferred that the PDW nematic domains were pinned by interactions with Zn impurity atoms at the Cu sites (Figure 3 (g,h)), revealing that this PDW state may be driven by strong-coupling physics and disorder effects.

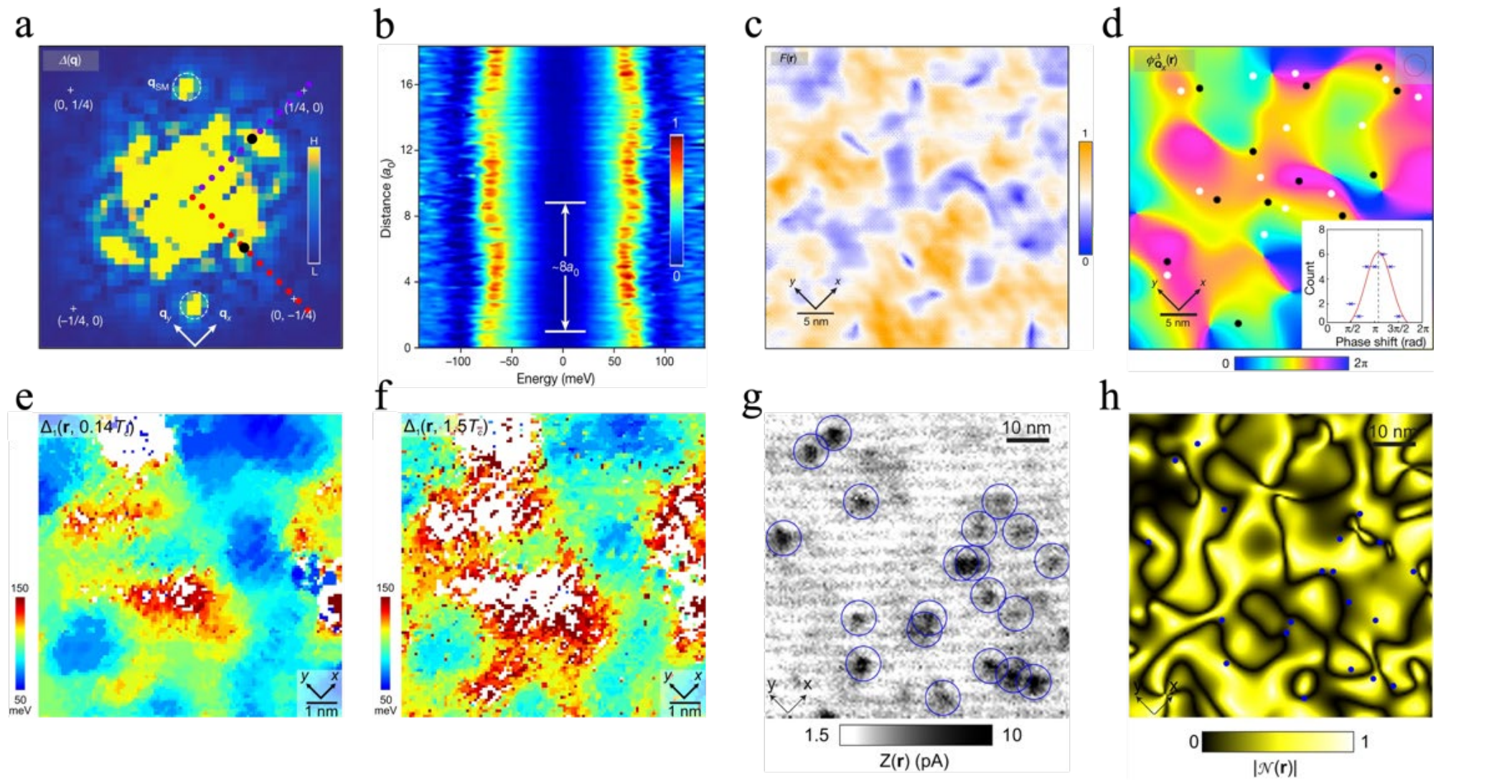

**Fig. 3** Analysis of SC energy gap map $\Delta(\boldsymbol{r})$ and 2D lock-in technique in Bi-2212. (**a-d**) Nematic PDW and phase analysis in near-optimally doped Bi-2212 [54]. (a) $\Delta(\boldsymbol{q})$, fourier spectrum of SC energy gap map. (b) Linecut of dI/dV spectra intensity. (c) The nematic order parameter map $F(\boldsymbol{r})$ defined by Formula (14). (d) $\Phi^{z}_{\boldsymbol{Q}_x}(\boldsymbol{r})$, the spatial phase of the PDW at wavevector $\boldsymbol{Q}_x$. The 2π topological defects in $\Phi^{z}_{2\boldsymbol{Q}_x}(\boldsymbol{r})$ (induced $2\boldsymbol{Q}_{2q-8a_0}$ charge order) are plotted on top of $\Phi^{z}_{\boldsymbol{Q}_x}(\boldsymbol{r})$. The inset shows the distribution of phase shift of $\Phi^{z}_{\boldsymbol{Q}_x}(\boldsymbol{r})$ values at all the locations where the 2π topological defects in $\Phi^{z}_{2\boldsymbol{Q}_x}(\boldsymbol{r})$ are found. (**e-f**) Energy gap modulations from superconductive to pseudogap phase in severely underdoped Bi-2212 [66]. (e) Measured $\Delta_1(\boldsymbol{r})$ at $T = 0.14\ T_c = 5$ K. (f) Measured $\Delta_1(\boldsymbol{r})$ at $T = 1.5\ T_c = 55$ K exhibits minimal changes compared to panel (e). (**g-h**) Zn-pinned PDW nematic domains in Bi-2212 [53]. (g) Zn impurities visualization through local maxima in $Z(\boldsymbol{r}) \equiv I(20\text{ mV}, \boldsymbol{r}) - I(-20\text{ mV}, \boldsymbol{r})$. The locations of Zn impurity atoms are shown in blue circles. (h) $N(\boldsymbol{r})$, the amplitude of the nematic order parameter, with the sites of Zn impurity resonances overlaid as blue dots. The author employed a distinct symbol $N(\boldsymbol{r})$ compared to $F(\boldsymbol{r})$ utilized in this review.

While previous experimental studies have primarily focused on underdoped or optimally doped cuprates, recent work by Wang *et al.* extended the study on PDW to the overdoped regime, specifically in $Bi_{2.08}Sr_{1.92}CuO_{6+\delta}$ (Bi-2201) [67]. In this system, they observed a robust, non-dispersive LDOS modulation with a periodicity of $4a_0/3$, accompanied by a clear π-phase shift between filled and empty states within the pseudogap energy window (20 – 55 meV), and this electronic state was attributed to a PDW order induced by the Amperean pairing with a finite momentum [67]. Moreover, domain-resolved analysis revealed locally unidirectional and short-range PDW textures, suggesting that quenched disorder and nematicity may pin the PDW, preventing long-range coherence.

Besides experimental studies, Choubey *et al*. employed a strong-coupling renormalized mean-field theory (RMFT) combined with Bogoliubov-de Gennes (BdG) equations to construct the atomic-scale electronic structure of a unidirectional PDW state with a periodicity of $8a_0$, coexisting with uniform $d$-wave superconductivity (DSC) [95]. By projecting the lattice Green's function onto the BiO surface using Wannier functions, they obtained theoretical local density of states $N(\boldsymbol{r}, E)$ and Bogoliubov quasiparticle scattering interference (BQPI), which were compared with high-resolution spectroscopic-imaging STM measurements. The close agreement in detail led the authors to propose a framework in which a PDW coexists with DSC in underdoped Bi-2212, with the pseudogap interpreted as the antinodal gap of the PDW state. Moreover, the PDW+DSC scenario accounts for the observed $4a_0$ CDW modulations as well as the critical point $p$*, associated with the disappearance of the PDW [95].

The STM studies of PDW in cuprates, especially in $Bi_2Sr_2CaCu_2O_{8+x}$, have provided crucial insights into the complex interplay between superconductivity and charge order, elucidating the physical mechanisms underlying the pseudogap phase and high-temperature superconductivity. Observations of PDWs coexisting with robust Cooper-pair condensates under varying conditions highlight their significance across the cuprate phase diagram and suggest new research avenues for exploring the impact of external perturbations on these intertwined states. These observations challenge existing theoretical frameworks and open up possibilities for manipulating superconducting properties in quantum technologies. Nevertheless, it is worth noting that the interpretation of the observed periodic modulations, particularly those with wavevectors corresponding to $4a_0$, $8a_0$ and $4a_0/3$ periodicities, remains an active area of debate, as Webb *et al.* demonstrated that the density-wave wavevector $\boldsymbol{Q}_{DW}$ undergoes a transition from commensurate to incommensurate with increasing local hole doping [96]. The precise distinction between the features induced by primary PDW and secondary features arising from other competing orders – such as CDWs, structural inhomogeneity, or quasiparticle interference – has not yet reached full consensus.

## 2.2 Iron-based superconductors

Similar to cuprates, the iron-based superconductors exhibit a complex phase diagram with a variety of symmetry-breaking electronic states, such as the nematic order and stripe antiferromagnetic order. Since their experimental discovery in 2008, iron-based superconductors have gained attention for their unconventional superconducting mechanisms, which are characterized by multi-band structures and strong electron correlations [97, 98]. The interplay between magnetism and superconductivity in these systems is particularly intricate, with orbital and spin fluctuations believed to mediate unconventional pairing states such as $s\pm$ wave [99]. Under in-plane magnetic fields, the anomalous upturn of the upper critical field at low temperatures was reported in iron-based superconductors, and was considered as a signature of the FFLO states [100, 101]. It indicates that PDW states may also exist in iron-based superconductors.

$A$EuFe$_4$As$_4$ ($A$=Rb, Cs) simultaneously hosts superconductivity and magnetism, exhibiting multigap superconductivity with a transition temperature of $T_c \approx 37$ K, occurring within the FeAs bilayers, while magnetic ordering emerges below $T_m \approx 15$ K due to the alignment of $Eu^{2+}$ spins, which induces ferromagnetic behavior [102]. By studying the cleaved $EuRbFe_4As_4$ (Eu-1144) single crystals, Zhao *et al.* reported the existence of a zero-field nematic PDW state, independent of other spatially ordered states [69]. Spatial modulations of the SC gap indicated the signatures of a PDW state with wavevector $\boldsymbol{Q}_{1q-8a_{\mathrm{Fe}}} = \frac{1}{8}\boldsymbol{Q}_{\mathrm{Bragg}}^{\mathrm{Fe}}$. The modulated gap followed a mean-field temperature dependence, vanishing with the loss of Eu-plane magnetism, and was further suppressed inside the magnetic vortices (Figure 4 (e-f)), indicating its coexistence with a uniform superconducting component. The PDW formation was attributed to exchange splitting of the Fermi surface, driven by Fe *d* and Eu *f* orbital coupling [103], with PDW order emerging at M point electron pockets and uniform superconductivity at Γ point hole pockets. Inter-pocket scattering was proposed to induce a PDW component across both pockets. However, the expected temperature-dependent $\boldsymbol{Q}$ vector, arising from decreasing spin splitting with rising temperature, was not observed experimentally, necessitating further investigation. Besides, this may also suggest that, under the proposed mechanism, the system lies in a strong-coupling regime where the temperature dependence of $\boldsymbol{Q}$ is too small to be detected. All these possibilities necessitate further investigation.

Fe(Te,Se) is notable for their enhanced superconductivity due to Se doping and the coexistence of topological and superconducting phases [17, 104]. The competition between antiferromagnetic and superconducting orders in Fe(Te,Se) makes it an ideal system for studying PDWs. Liu *et al.* reported the signatures of PDW states in epitaxially grown Fe(Te,Se) films at domain walls [61]. Although no spatial modulations were observed in the topographic images or their FFTs, measurements of zero-bias conductance (ZBC), coherence peak height and SC energy gap revealed modulations with wavevector $\boldsymbol{Q}_{1q-a_{\mathrm{Fe}}/0.28} = 0.28\boldsymbol{Q}_{\mathrm{Bragg}}^{\mathrm{Fe}}$ exclusively within the domain walls, supporting the existence of a PDW order, as shown in Figure 4 (a-c). Phase analysis revealed that the predicted π-phase shift in the PDW order at topological defects led to a vortex-like 2π phase-winding in the induced $2\boldsymbol{Q}_{1q-a_{\mathrm{Fe}}/0.28}$ CDW (Figure 4 (d)). The emergence of the quasi-one-dimensional PDW state was attributed to local lattice distortions at domain walls that break spatial symmetries, enhancing spin-orbit coupling (SOC). Rashba and Dresselhaus SOC split the domain wall states in momentum space and result in two spin-mixed bands, leading to a quasi-one-dimensional equal-spin pairing PDW confined to the walls. Theoretically, Zhang and Wang have proposed a spinful Kitaev chain with nearest-neighbor pairing and generic spin-orbit coupling, predicting the emergence of a spin-triplet PDW state characterized by a spatially modulated d-vector and topological superconductivity with Majorana zero modes at the ends. This model offers insights into the quasi-one-dimensional PDW observed at domain walls in Fe(Te,Se) and highlights the interplay between SOC, pairing interactions, and chemical potential in stabilizing exotic superconducting phases [105].

Although local lattice distortions at domain walls facilitate the formation of PDW, observations indicate that PDW states can also occur on the atomically flat surface. Wei *et al.* continued their investigation on $FeTe_{1-x}Se_x$ films (x ≈ 0.7) epitaxially grown on STO substrates, focusing on the atomically flat surface [68]. The commensurate CDW with wavevector $\boldsymbol{Q}_{2q-4a_{\mathrm{Fe}}} = \frac{1}{4}\boldsymbol{Q}^{\mathrm{Fe}}_{\mathrm{Bragg}}$ in this system broke $C_4$ rotational symmetry and exhibited stripe order features. Further analysis of coherence peak height (CPH) revealed $4a_{\mathrm{Fe}}$ periodic modulations along two orthogonal directions with anisotropic intensity, indicating the presence of a rotational symmetry breaking PDW state. Moreover, the phase features of the CDW, including phase slip domain walls and 2π phase-winding vortices, were mirrored in the PDW phase, reinforcing the connection between the smectic CDW and the induced PDW. The directional behavior of the CDW, dependent on energy, can be diminished by magnetic fields or local defects, hinting at a mechanism driven by the coupling between CDW and PDW order parameters.

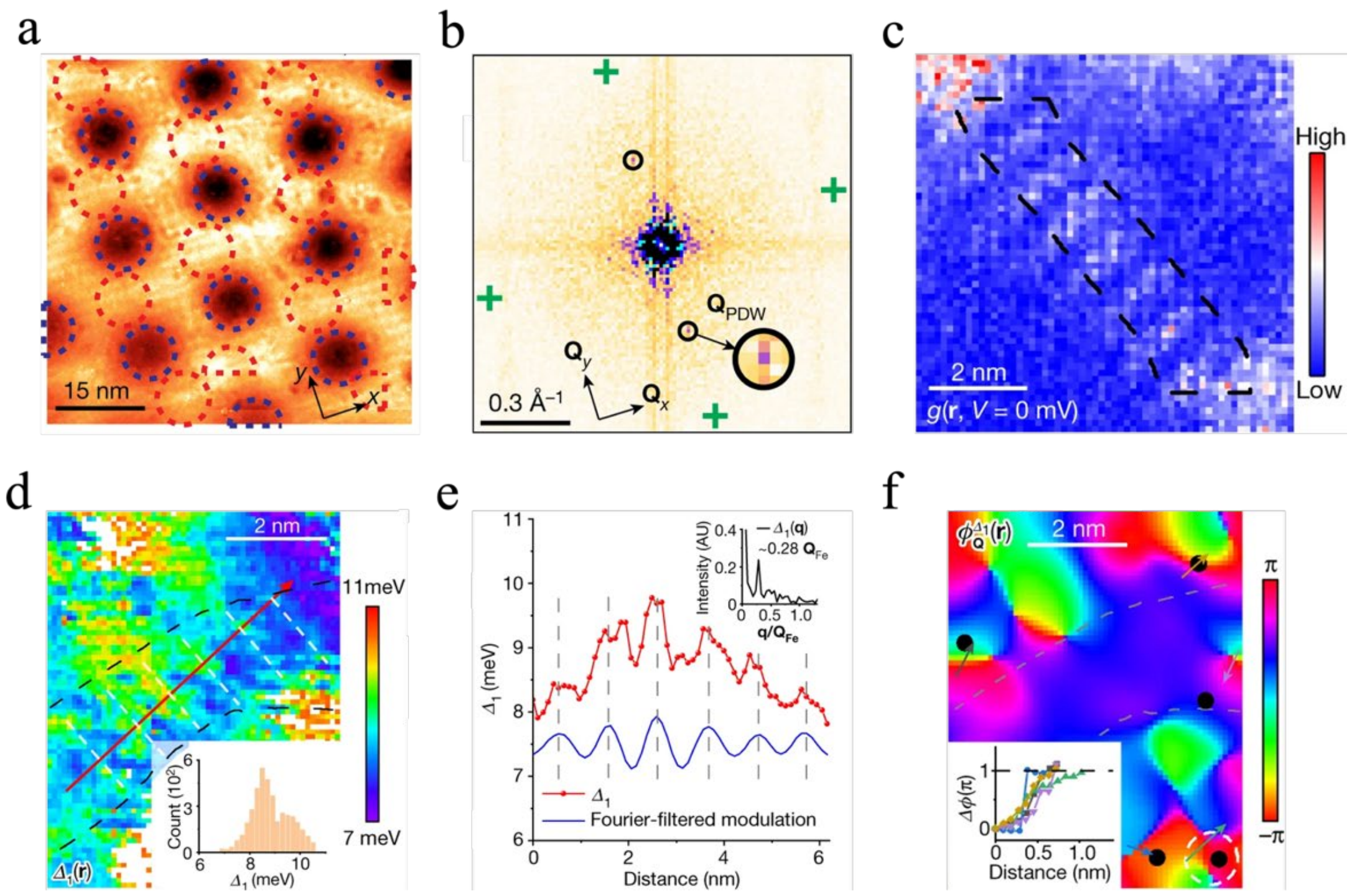

**Fig. 4** Signatures of PDW in Iron-based superconductors. (**a-b**) The existence of a zero-field PDW state in the $EuRbFe_4As_4$ single crystals [69]. (a) Map of $N(\boldsymbol{r}, E \approx \Delta_0)$ at 9 T, displaying a hexagonal vortex lattice; blue circles highlight the vortex halos, while red circles highlight their uniform translation outside these regions. (b) FT of panel (a), with the PDW wavevector $\boldsymbol{Q}_{1q-8a_{\mathrm{Fe}}}$ marked by black circles. (**c-f**) Detection of PDW at domain walls in Fe(Te,Se) films [61]. (c) ZBC map with the domain wall highlighted by black dashed lines. (d) Spatial map of $\Delta(\boldsymbol{r})$. (e) The red curve represents the linecut of $\Delta(\boldsymbol{r})$ along the trajectory indicated as the red arrow in panel (d). The blue curve represents the extracted gap modulation by applying Fourier filter. The inset is FT of the linecut. (f) Spatial phase variation of the gap energy modulation at wavevector $\boldsymbol{Q}_{1q-a_{\mathrm{Fe}}/0.28}$, measured in the same domain wall as in panel (d). Inset: phase profile along the arrows, along which π-phase shifts of the PDW occur near the CDW vortices marked by blacked dots.

The PDW states discussed above break the long-range lattice translational symmetry, while pair density modulation (PDM), which only breaks intra-unit-cell symmetries of the space group has been proposed by Wei *et al* [7] and Kong *et al.* [8]. As discussed previously in this review, PDM can be regarded as a type of PDW in a broad sense. Wei *et al.* reported the PDM in the Fe(Te,Se) films epitaxially grown on the STO substrates and found that the Fe atoms within a unit cell were equivalent, while the Te/Se atoms were inequivalent due to the glide-mirror symmetry breaking introduced by the STO substrate [7]. The energy gaps $\Delta_1(\boldsymbol{r})$ and $\Delta_2(\boldsymbol{r})$, along with the coherence peak sharpness, exhibited spatial modulation at Bragg vector $\boldsymbol{Q}_{2\mathrm{q}-a_{\mathrm{Te,Se}}} = \boldsymbol{Q}_{\mathrm{Bragg}}^{\mathrm{Te,Se}}$ (Figure 5 (a,b)). Additionally, the local maxima and minima of the modulation aligned respectively with the crystallographic positions of the uppermost and lowermost Te/Se atoms.

In addition to the Fe(Te,Se) films, the PDM has been also observed in exfoliated $FeTe_{0.55}Se_{0.45}$ thin flakes [8], where the coherence peak modulation aligned with the crystal lattice, and the SC energy gap Δ exhibited unexpectedly strong modulations at Bragg vectors $\boldsymbol{Q}_{2\mathrm{q}-a_{\mathrm{Fe}}} = \boldsymbol{Q}_{\mathrm{Bragg}}^{\mathrm{Fe}}$ (Figure 5 (c,d)). The FT filtered image at $\boldsymbol{Q}_{2\mathrm{q}-a_{\mathrm{Fe}}}$ revealed a strong gap map distortion dividing the region into small domains, as shown in Figure 5 (e). Within X-Domain, the gap maxima appeared at iron sublattice $Fe_x$, and the gap minima with the other $Fe_y$. Conversely, this corresponds to the Y-domain. The PDM state may originate from a synergy between broken glide-mirror symmetry and a distinct nematic distortion with a director aligned along the next-nearest-neighbor direction, as shown in Figure 5 (f). While the glide-mirror symmetry breaking leads to inequivalent next-nearest-neighbor hopping amplitudes between the two iron sublattices, the rotated nematicity further lifts their electronic degeneracy. This work provides direct observation of a PDM state in an iron-based superconductor, revealing a novel class of SC order.

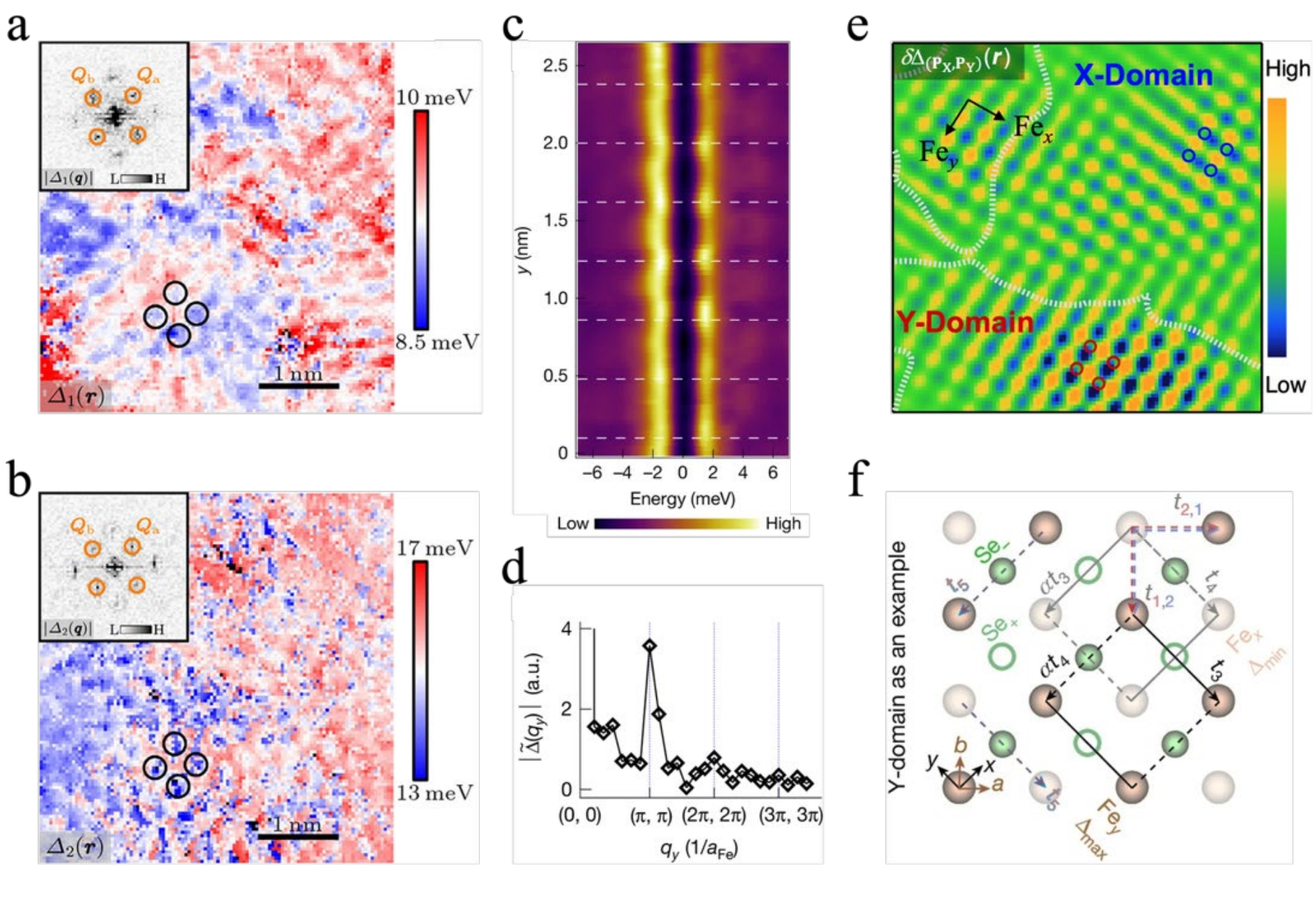

**Fig. 5** Signatures of PDM in Iron-based superconductors. (**a-b**) PDM state in Fe(Te,Se) films [7]. (a) SC gap map of $\Delta_1(\boldsymbol{r})$. Inset: FT of $\Delta_1(\boldsymbol{r})$. (b) SC gap map of $\Delta_2(\boldsymbol{r})$. Inset: FT of $\Delta_2(\boldsymbol{r})$. The PDM wavevector, which coincide with the Bragg wavevectors of the top-layer Te/Se sublattice, are indicated by orange circles. (**c-f**) PDM state in $FeTe_{0.55}Se_{0.45}$ thin flakes [8]. (c) Linecut of d$I$/d$V$ spectra intensity. The SC gap is minimized at Se+ atom sites which are marked by dashed white lines. (d) The FT spectra of the SC gap modulation along the $y$-axis reveals a modulation wavevector corresponding to the Bragg vector. (e) FT filtered image at $\boldsymbol{Q}_{2q-a_{\mathrm{Fe}}}$ of $\delta\Delta(\boldsymbol{r})$. The circles mark the positions of Se+ atoms. (f) The model proposed to explain the PDM: $t_1$ is the nearest neighbor hopping integral and $t_{2,3}$ are the next-nearest neighbor hopping integrals. The nematic distortion generated a parameter $\alpha$, resulting in the stripe pattern.

Furthermore, Zhang *et al.* studied epitaxially grown single-layer FeSe on the STO substrate, where the interfacial structure broke inversion and $C_4$ rotational symmetries [70]. Their STS analysis revealed periodic spatial modulation between Se and Fe sublattices, as well as a negative correlation between the superconducting gap and coherence peak height. They also observed spatial modulation with a $\sqrt{2}a_{\mathrm{Fe}}$ period in Cooper-pair density, attributing it to *p-d* orbital hybridization, which alters local pairing strength and superfluid density. Inter-unit-cell (inter-UC) PDW with a period of $5.4a_{\mathrm{Fe}}$ appears near domain walls in a narrow energy range [61] around $\Delta^*$, which represents a pair of kinks and originates from quasiparticle excitations, while intra-unit-cell (intra-UC) PDW with a period of $\sqrt{2}a_{\mathrm{Fe}}$ extends over a wider energy range, linked to multiple symmetry breakings and spin-orbit coupling at the interface. In practical, by combining this work with results observed in PDM, one may infer that the inter-UC PDW in FeSe/STO may be related to the PDM state.

The microscopic mechanism underlying PDW formation is still unclear. While spin-orbit coupling, inter-orbital hybridization, and local symmetry breaking have been proposed as contributing factors in specific systems such as Fe(Te,Se), a unified theoretical framework applicable across different materials is still lacking. Furthermore, the interplay between magnetism and superconductivity in materials such as $EuRbFe_4As_4$ complicates the understanding of PDW formation and its coexistence with uniform superconductivity. However, the experimental advances using STM/STS discussed above are thought to underscore the intricate interplay of lattice, spin, and orbital effects in shaping PDW states in iron-based superconductors and to offer new insights into unconventional pairing mechanisms.

### 2.3 Spin-triplet Superconductors

In spin-singlet superconductors, the orbital wave function of Cooper pairs is symmetric about the origin and the spin orientation is antisymmetric. Consequently, the two electrons must form a spin-singlet state [106]. In contrast, if the orbital wave function is antisymmetric about the origin with nodes, the Cooper pair must be in the spin triplet state, resulting in the odd order parameter: $\Delta(\boldsymbol{k}) = -\Delta(-\boldsymbol{k})$ [107]. The intriguing phenomenon has been demonstrated in superfluid $^3$He [108] and is believed to occur in strongly correlated electron superconductors [109, 110].

$UTe_2$ is the newly discovered spin-triplet superconductor because of the coexistence of magnetism and superconductivity, supported by extensive experimental studies [106, 111–113]. It exhibits a superconducting transition temperature of $T_c \approx 1.6$ K [106, 114] and large upper critical field beyond Pauli limit [115]. Numerous theoretical and experimental studies have shown intriguing properties in $UTe_2$, indicating the presence of novel states [116, 117]. Additionally, various studies have reported the signatures of PDW states in $UTe_2$.

Aishwarya *et al.* revealed multi-component, incommensurate CDWs in $UTe_2$ through STM imaging, which were found to be sensitive to the magnetic field [118]. When subjected to a magnetic field, the CDWs were significantly suppressed, exhibited mirror symmetry breaking (Figure 6 (a)), and eventually vanished at the upper critical field $H_{c2}$, suggesting that the CDW and superconductivity not only coexist but are also strongly coupled. Theoretical analysis attributes this behavior to the presence of triplet PDW order, which leads to a direction-dependent critical field [111], and the two mirror-symmetry-related CDWs are suppressed to different extents when mirror symmetry is broken by the external magnetic field.

Almost at the same time, Gu *et al.* reported the signatures of PDW in $UTe_2$ [23]. The spatial modulation of coherence peak, Andreev resonances, and SC energy gap revealed incommensurate PDW order with wavevectors $\boldsymbol{P}_{1,2,3}$ with a characteristic energy scale of 10 μeV for peak-to-peak modulations, and those wavevectors were consistent with those of CDW order they observed (Figure 6 (b-d)). Additionally, they observed that the CDW image in the non-superconducting state and the PDW image in the superconducting state were spatially registered with high precision, but exhibited a relative phase shift of π, revealing the anti-correlation between PDW and CDW, as shown in Figure 6 (e). Based on these experimental observations, they propose that the PDW in this system may be induced by the CDW and superconductivity and is likely a spin-triplet PDW which is unprecedented for superconductors.

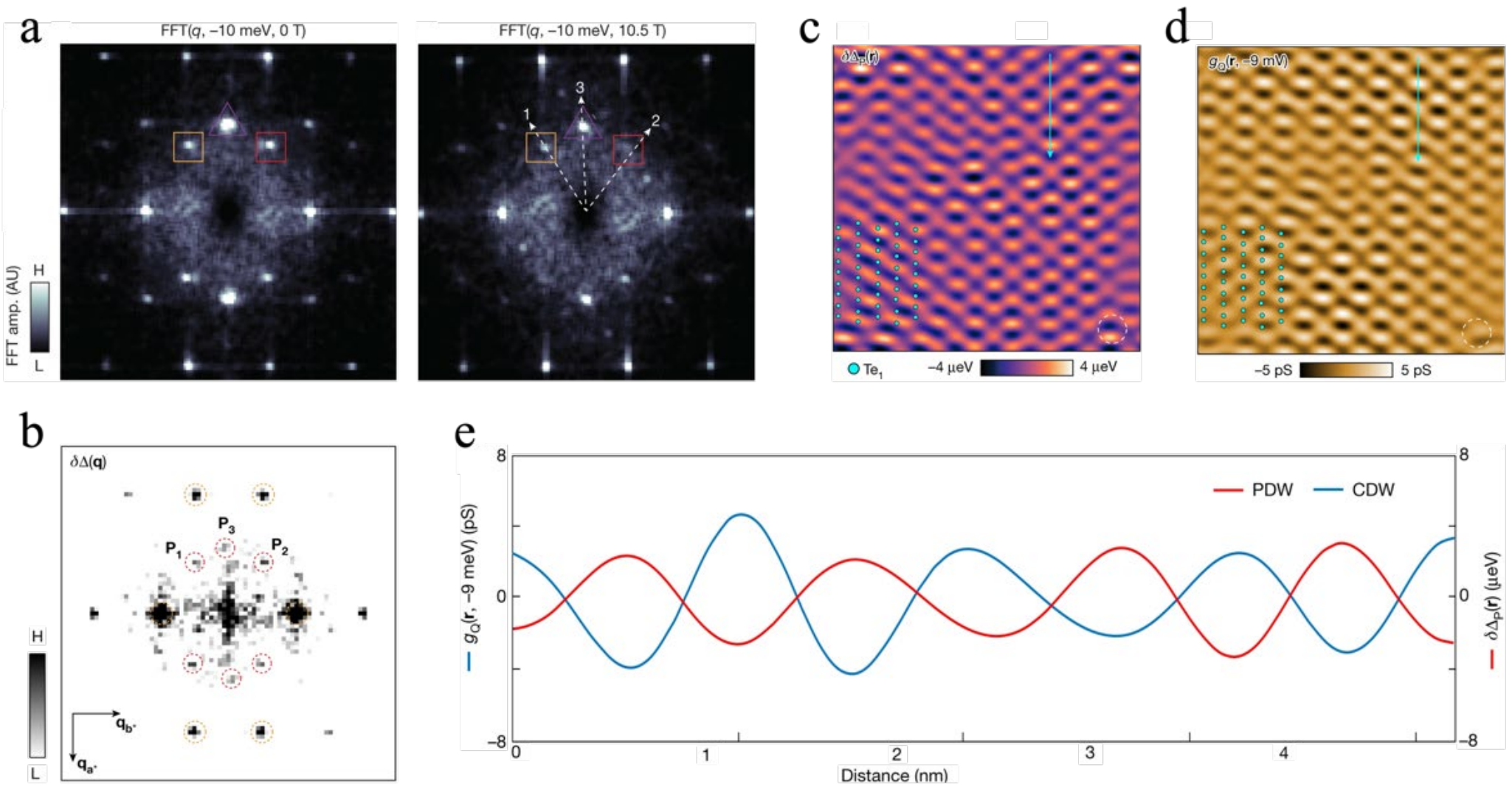

**Fig. 6** The magnetic-field-sensitive CDW and the interplay between CDW and PDW in $UTe_2$. (**a**) FFTs of LDOS maps at 0 T and 10.5 T of $UTe_2$ single crystals, with the CDWs marked in orange and red squares, and the purple triangle, showing the magnetic field response of the CDWs in $UTe_2$ [118]. (**b-e**) signatures of PDW and the interplay between PDW and CDW in $UTe_2$ single crystals [23]. (b) FT of $\delta\Delta(\boldsymbol{r})$, which reveals three PDW peaks at $\boldsymbol{P}_{1,2,3}$ marked by dashed orange circles. (c) Inverse Fourier transform of filtered $\delta\Delta(\boldsymbol{q})$ at $\boldsymbol{P}_{1,2,3}$ visualizes the PDW. (d) Inverse Fourier transform filtered $g(\boldsymbol{q}, -9\ \text{mV})$ at $\boldsymbol{Q}_{1,2,3}$ depicts the CDW. (e) Profile of $g_Q(\boldsymbol{r}, -9\ \text{mV})$ and $\delta\Delta_P(\boldsymbol{r})$ along the same trajectory in the same FOV.

Additionally, LaFleur *et al.* demonstrated that the CDW persists well above the superconducting transition temperature and can still be gradually suppressed by magnetic fields, indicating that the CDW is not induced by superconductivity but may originate from an underlying spin mechanism in $UTe_2$. Moreover, the spatial evolution of CDW domains exhibits strong reversibility upon thermal or magnetic-field cycling, highlighting their sensitivity to external tuning and the potential role of static disorder in stabilizing the short-range order [119]. Recent STM measurements by Talavera *et al.* suggested that the CDW in $UTe_2$ possesses primitive wavevectors residing entirely within the surface Brillouin zone [120]. Besides, Theuss *et al.* observed a smooth variation in the elastic modulus across a temperature range from 2 K to 280 K, without detecting any sharp changes or discontinuities indicative of electronic phase transitions. This observation suggests the absence of CDW and PDW in the bulk phase of $UTe_2$. The lack of a thermodynamic phase transition in the bulk implies that the CDW and PDW observed in STM studies may be confined to the surface [121], necessitating further investigation.

## 2.4 Kagome Superconductors

Recently, transition-metal kagome materials have garnered attention for their intriguing metallic-phase properties, including geometric frustration, flat bands, Dirac fermion crossings, and van Hove singularities [122, 123]. These features make kagome lattices ideal for studying the complex interactions between electronic geometry, topology, and correlations [124, 125].

Chen *et al.* observed V-shaped superconducting gap and PDW order in the recently-discovered $CsV_3Sb_5$ [71]. Within the V-shaped superconducting gap, they proposed that the non-zero LDOS may originate from localized vortex-antivortex core states and itinerant nodal quasiparticles. In addition to the CDW orders with wavevectors $\boldsymbol{Q}_{1q\text{-}4a}$ and $\boldsymbol{Q}_{3q\text{-}2a}$ observed in both topography and d*I*/d*V* maps, several superconductivity-related physical quantities, including the energy gap $\Delta(\boldsymbol{r})$, coherence peak height $P(\boldsymbol{r})$, zero-bias conductance $G_0(\boldsymbol{r})$, and gap depth $H(\boldsymbol{r}) = P(\boldsymbol{r}) - G_0(\boldsymbol{r})$, exhibited pronounced spatial modulations characterized by a hexagonal wavevector $\boldsymbol{Q}_{3q-4a/3} = \frac{3}{4}\boldsymbol{Q}^{a,b}_{\text{Bragg}}$, as shown in Figure 7 (a-c). These modulations suggest the signatures of a three-component (3Q) PDW that remains nondispersive near the superconducting energy gap. Moreover, $P(\boldsymbol{r})$ and $G_0(\boldsymbol{r})$ displayed out-of-phase modulations, while $P(\boldsymbol{r})$ and $H(\boldsymbol{r})$ exhibited beating patterns from the leading $4a_0/3$ PDW and a weaker $2a_0$ CDW, as shown in Figure 7 (d). These observations revealed a novel bidirectional electronic density wave that weakly perturbs the total charge density but strongly modulates superconducting coherence. The authors proposed a roton PDW mechanism, wherein the observed PDW with proposed complex order parameter behaves like a roton excitation. This intertwined quantum state, composed of delocalized Cooper pair modulations and localized charge or vortex–antivortex excitations, breaks time-reversal symmetry and could account for the observed low-energy spectral features and spatially modulated superconducting properties. Magnetic field-dependent measurements revealed that the 3Q PDW at $\boldsymbol{Q}_{3\mathrm{q}-4a/3}$ survives in vortex halos and persists even after superconductivity is suppressed, indicating a pseudogap phase. This PDW remained confined to the low-energy pseudogap range and was detectable even in the normal state at 4.2 K (Figure 7 (e)), suggesting that it may play a fundamental role in the emergence of the intertwined electronic order and the formation of the pseudogap.

Unconventional chiral charge order has been experimentally observed in kagome materials, drawing increasing interest due to its intertwining electronic chirality [126, 127]. Deng *et al.* claimed they observed chiral superconductivity and PDW modulations in $KV_3Sb_5$ and $CsV_3Sb_5$ [72]. Spatial $2a \times 2a$ modulations of the SC gap and pairing density, along with corresponding FFT signatures, indicated a PDW with wavevector $\boldsymbol{Q}_{3\mathrm{q}-2a} = \frac{1}{2}\boldsymbol{Q}_{\mathrm{Bragg}}$, consistent with the underlying charge order. Notably, the FFT of the SC gap, $\Delta(\boldsymbol{q})$, exhibits a distinct chiral character: the modulation intensity increases in a clockwise direction (Figure 7(f)), thereby demonstrating the chirality of the PDW. The residual Fermi arcs associated with *d*-orbital reconstruction were observed in their QPI measurements, which were interpreted to be interorbital *p*-*d* PDW pairing that gaps part of the Fermi surface, as shown in Figure 7(g,h).

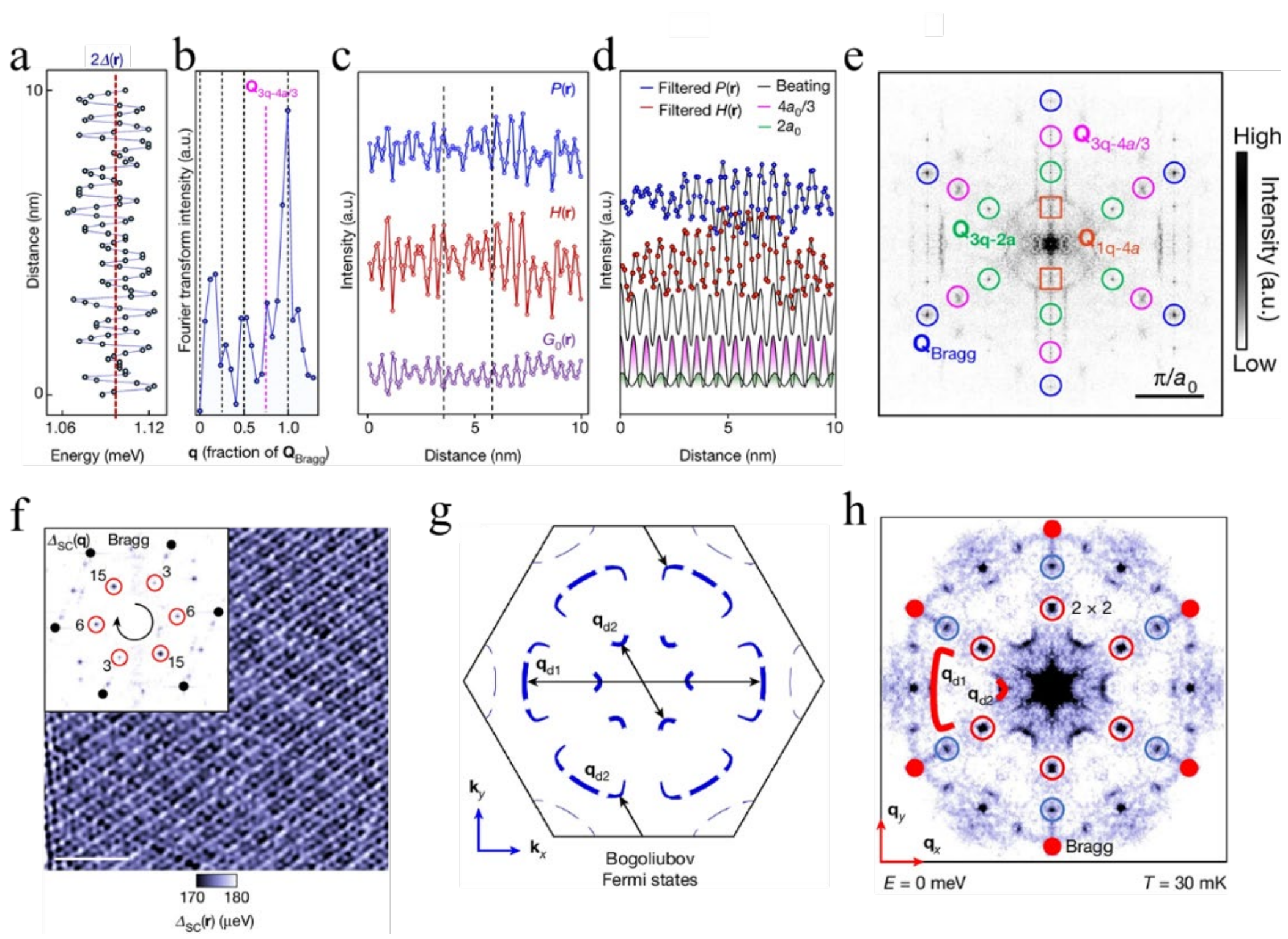

**Fig. 7** Signatures of PDW in Kagome lattice superconductors. (**a-e**) Roton PDW state in $CsV_3Sb_5$ [71]. (a) Linecut of $2\Delta(\boldsymbol{r})$ clearly shows the spatial modulation. (b) The FT spectrum of (a) shows the spatial modulation at wavevectors $\boldsymbol{Q}_{3q\text{-}4a/3}$, highlighted by the pink dashed line. (c) Spatial modulations of $P(\boldsymbol{r})$, $G_0(\boldsymbol{r})$, and $H(\boldsymbol{r})$. (d) The Upper two curves are the Bragg-filtered modulations of $P(\boldsymbol{r})$ and $H(\boldsymbol{r})$, while the bottom three curves represent the beating patterns simulated using $\boldsymbol{Q}_{3q\text{-}2a}$ (green curve) and $\boldsymbol{Q}_{3q\text{-}4a/3}$ (pink curve). (e) The magnitude of the drift-corrected, two-fold symmetrized FFT of d$I$/d$V$ ($\boldsymbol{r}$, −5 mV) map at the vortex halo in zero magnetic field at temperature 4.2 K. (**f-h**) Chiral PDW and residual Fermi arcs in $KV_3Sb_5$ [72]. (f) The SC energy gap map $\Delta(\boldsymbol{r})$. The inset is the corresponding FFT spectra. (g) Schematic of arc-like Bogoliubov Fermi states under a $p$ – $d$ orbital-selective PDW combined with uniform $p$-orbital pairing. Scattering among these states produces arc-like QPI signals at $\boldsymbol{q}_{d1}$ and $\boldsymbol{q}_{d2}$. (h) Symmetrized QPI at zero energy. The Bragg vectors, the $\boldsymbol{Q}_{\mathrm{Bragg}}/2$ and the $3\boldsymbol{Q}_{\mathrm{Bragg}}/4$ wavevectors are marked by the red points, red circles and the blue circles respectively.

In addition to the bulk PDW order, Han *et al.* reported a novel quasi-2D SC state, along with a PDW, on both spontaneously formed and atomically tailored 2×2 Cs ordered superlattice surfaces of the kagome metal $CsV_3Sb_5$ [73]. This quasi-2D superconductivity, emerging from the quasi-2D surface states, is characterized by the larger energy gap, higher upper critical field, higher critical temperature, and smaller coherence length compared to its bulk counterpart [71, 128, 129], and is capable of hosting intertwined density wave states that are absent in bulk superconductivity. Spectroscopic d$I$/d$V$ maps and corresponding FFT analysis revealed a novel, nondispersive 4×4 spatial modulation – twice the periodicity of the 2×2 Cs order – that persisted across the energy range linked to quasi-2D superconductivity (Figure 8 (a-c)), distinguishing it from quasiparticle interference effects. Furthermore, spatial variations in the quasi-2D SC gap $\Delta_{2D}$ exhibited $2a_0$ and $4a_0$ periodicities, indicating the formation of PDW.

The experimental observations promote the study of theoretical models of PDWs in kagome superconductors. Jin *et al.* proposed that the nesting of the Fermi Surface results in a time reversal and inversion symmetry breaking PDW state with stable Bogoliubov Fermi pockets [130]. Zhou and Wang presented significant advances in the understanding of quantum states within kagome lattice superconductors, specifically in $AV_3Sb_5$ ($A$ = K, Rb, Cs) [35]. They elucidated the mechanism of the metallic CDW state, characterized by circulating loop currents and evolving into a doped orbital Chern insulator near van Hove filling. This results in Chern Fermi pockets (CFPs) with concentrated Berry curvature and orbital magnetic moments, providing an explanation for the observed quantum oscillations and a significant intrinsic anomalous Hall effect (AHE). Finite momentum pairing on these CFPs gives rise to novel roton PDW states, featuring a vortex-antivortex lattice, as shown in Figure 8 (d,e). This PDW state, characterized by a 3Q structure with periodicity that modulates the conventional superconducting gap and coherence, contributes to novel charge-4*e* and charge-6*e* superconductivity through a unique mechanism of staged melting of this lattice and further provides plausible explanation for the $Q_{3q\text{-}4a/3}$ PDW observed in $CsV_3Sb_5$ [22, 71]. The PDW observed with STM experimentally [73, 71, 72] can be explained well with their theoretical work [35]. Besides, Yao *et al.* also identified the presence of $2a_0\times2a_0$ 3Q PDW states near the upper van Hove singularity in the kagome lattice theoretically [131], classifying them into chiral and non-chiral types, with the chiral states exhibiting topological properties independent of spin-orbit coupling. The chiral 3Q PDW was found to achieve a fully gapped state through higher-order scattering processes and was predicted to give rise to quantized thermal Hall conductance alongside anisotropic superconducting gaps. They also analyzed the strong coupling between PDW and CDW, which potentially explains experimental phenomena in $AV_3Sb_5$, such as residual Fermi surfaces and fractional quantum flux under a chiral CDW background.

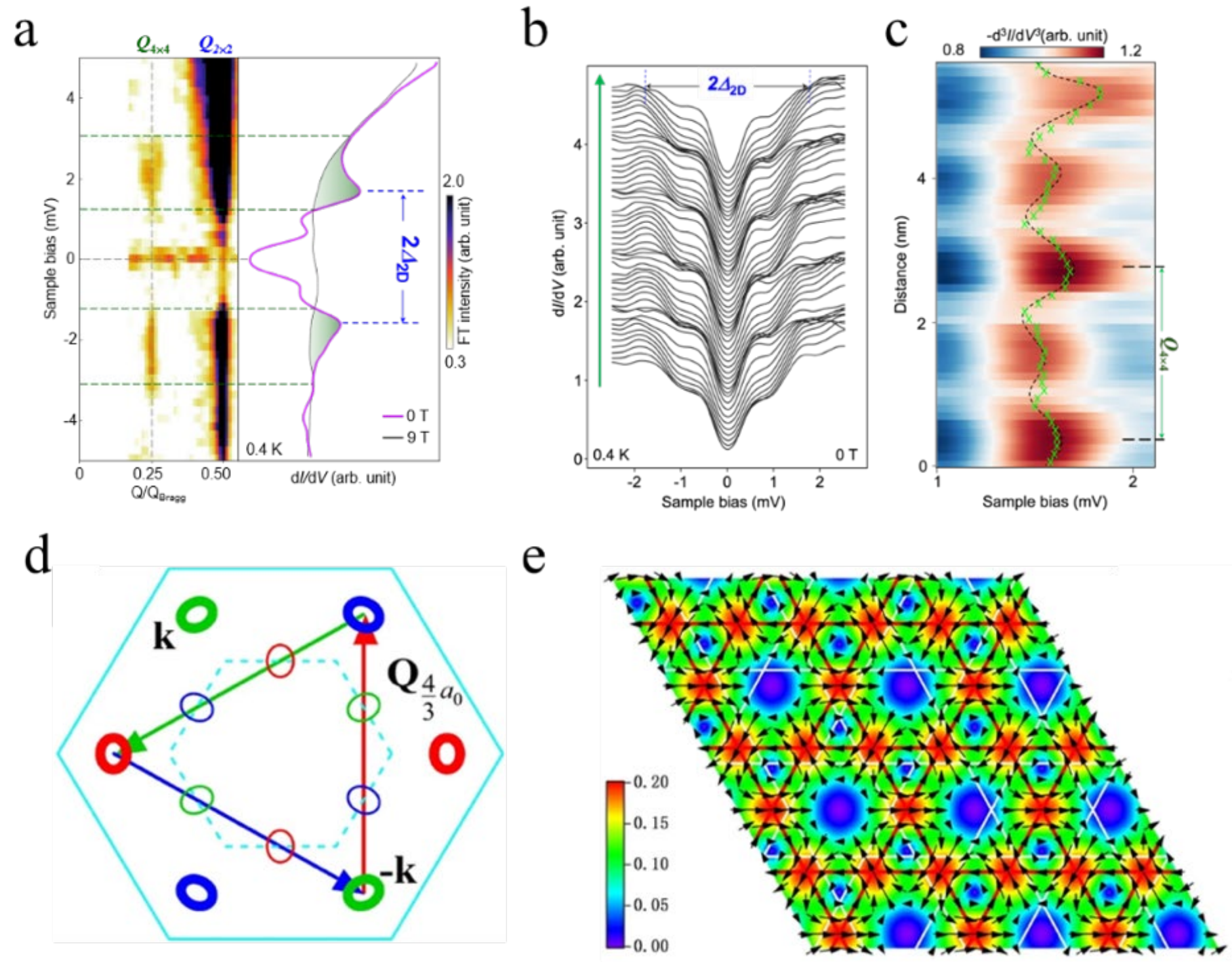

**Fig. 8** The 4×4 PDW in $CsV_3Sb_5$ and theoretical framework for the PDW in kagome superconductors. (**a-c**) Observation of the 4×4 pair density wave modulation of $\Delta_{2D}$ at the 2×2 Cs ordered surface of $CsV_3Sb_5$ [73]. (a) Left panel: FT linecut of d$I$/d$V$ maps. Right panel: d$I$/d$V$ spectra taken at 0 T and 9 T. (b) The waterfall plot of linecut of d$I$/d$V$. (c) Zoomed-in intensity plot of the -$d^3I/dV^3$ linecut corresponding to panel (b). (**d-e**) 3Q PDW and gapped Chern Fermi pockets [35]. (d) Schematic illustration of the CFPs. These pockets are connected by characteristic PDW momenta, enabling the formation of finite-momentum pairing. (e) Real-space profile of the complex PDW order parameter with $4/3a_0 \times 4/3a_0$ periodicity. Color intensity encodes the pairing amplitude, while arrows indicate the local phase. The red mesh outlines the emergent PDW Kagome pattern superimposed on the underlying Kagome lattice (white).

## 2.5 TMDs Superconductors

Transition metal dichalcogenides (TMDs), composed of transition metal elements and chalcogens, possess a distinctive layered structure analogous to that of graphene, where adjacent layers are held together by van der Waals forces [132, 133]. These materials exhibit intriguing physical properties, including opto-valleytronic properties [134, 135], tunable band gap [136], superconductivity [137–143], topological properties [144], optical properties [145], among others. Such properties significantly enhance their potential in various device applications [146]. The PDW states in TMDs have been proposed to arise through several mechanisms [147, 148]. Despite these predictions, experimental observation of the PDW states in TMDs remains elusive.

### 2.5.1 $NbSe_2$

2H-$NbSe_2$, a typical Ising superconductor, has been extensively studied both experimentally and theoretically for its superconductivity and CDW states [149, 150], which exhibit weak coupling between the superconducting and CDW order parameters [143]. However, Liu *et al.* discovered the PDW states originating from the coupling between CDW and superconductivity in $NbSe_2$ crystals [74]. The electron-pair density $N_{cp}(\boldsymbol{r})$ (Figure 9 (a)), obtained from Formula (8), and the energy gap $\Delta(\boldsymbol{r})$ revealed spatial modulations at the same wavevector $\boldsymbol{Q}_{3q-3a} = \frac{1}{3}\boldsymbol{Q}_{\mathrm{Bragg}}$ as the CDW, as revealed by $N_Q(\boldsymbol{r}) = g(\boldsymbol{r}, V)$. Moreover, the quasiparticle density $N_Q(\boldsymbol{r})$ and electron-pair density $N_{cp}(\boldsymbol{r})$, measured on a vortex core, showed that the combined PDW amplitude decayed consistently with the background superfluid density $N_s(\boldsymbol{r}) = N_{cp}(\boldsymbol{r}) - N_p(\boldsymbol{r})$, as expected from Ginzburg-Landau theory for a CDW-induced PDW state, where $N_P(\boldsymbol{r})$ denotes the filtered PDW map obtained through Formula (10, 11), and $N_c(\boldsymbol{r})$ denotes the filtered CDW map. Further analysis revealed that $N_c$ and $N_P$ were phase-shifted by $2\pi/3$, which resulted from the electron-pair wavefunction's $\boldsymbol{k}$-space structure factor. In $NbSe_2$, lattice-locked 3×3 commensurate CDW domains were separated by discommensurate regions where the CDW phase shifted by $\pm 2\pi/3$, as shown in Figure 9 (b, c). These phase shifts were observed in both the CDW and PDW states at nearly identical locations, suggesting that the PDW domains may be influenced by the preexisting CDW, indicating a CDW-superconductivity coupling mechanism.

Additionally, Cao *et al.* investigated PDW in $NbSe_2$ [75] and revealed its sensitivity to distinct CDW regions. Their study identified two distinct CDW regions – hollow-centered-CDW (HC-CDW) and anion-centered-CDW (AC-CDW) – consistent with previous research on CDW states of $NbSe_2$ [151]. PDW modulations exhibited $C_3$-symmetry breaking in the AC-CDW regions, while they remained commensurate with the CDW in the HC-CDW regions, as shown in Figure 9 (d-f). These observations suggest that the interactions between PDW and CDW cause spatial variations in the superconducting order parameter, thereby inducing anisotropy in the electron pairing density. The authors proposed that this nematic phase may be associated with a mixture of different superconducting pairing channels or a strong coupling effect of the CDW on the superconducting phase [141, 152]. The structural characteristics of the AC-CDW may locally induce more complex electronic interactions, whereas the HC-CDW likely maintains a more regular distribution and interaction of electrons. These results are consistent with previous work on $NbSe_2$ [74], particularly their focus on HC-CDW regions.

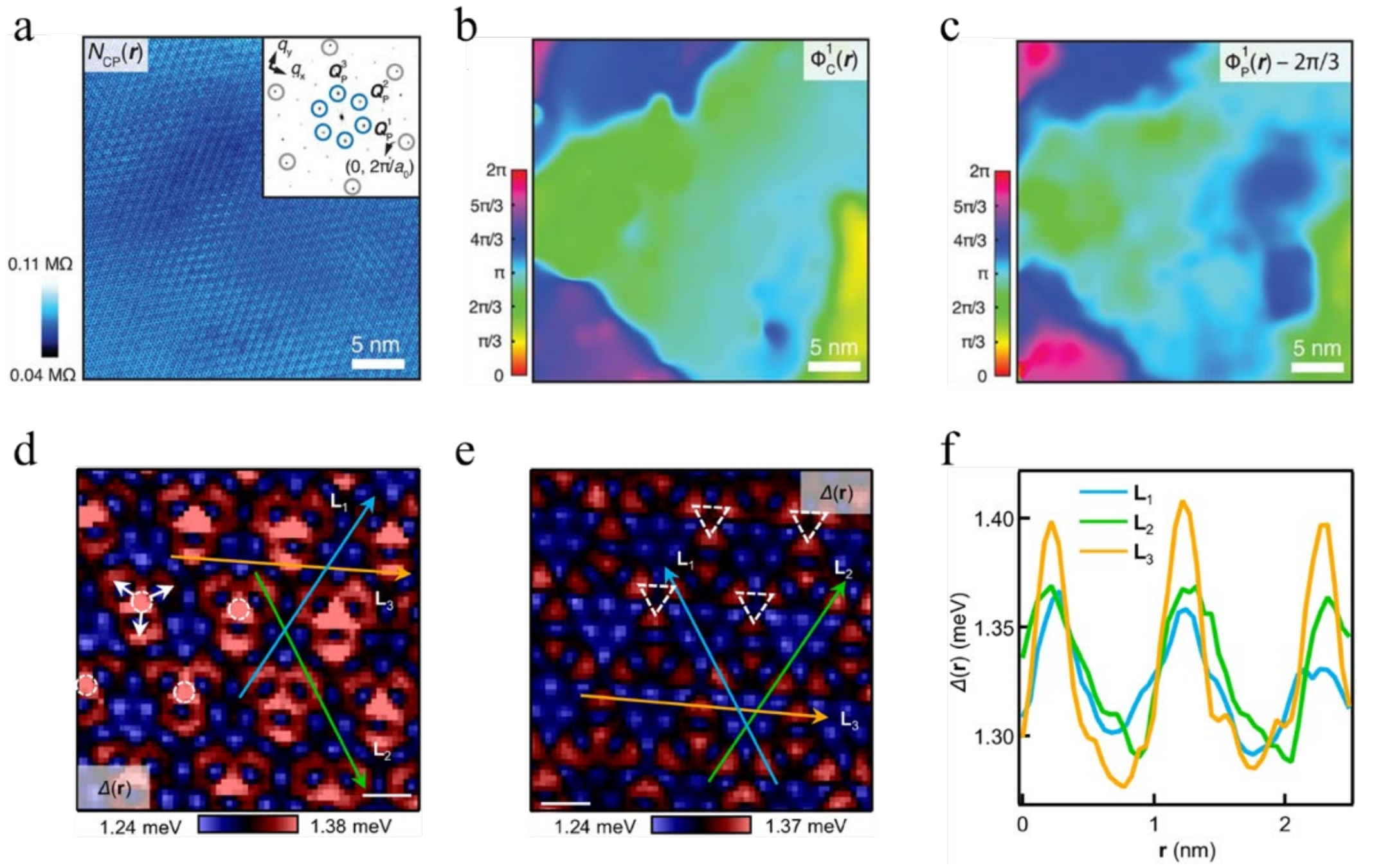

**Fig. 9** Signatures of PDW in $NbSe_2$ crystals. (**a-c**) The CDW-induced PDW states in $NbSe_2$ single crystals [74]. (a) Measured electron-pair density $N_{cp}(\boldsymbol{r})$. The inset is the Fourier spectrum with PDW wavevectors $\boldsymbol{Q}_{3q-3a}$ marked within blue circles. (b) Measured CDW spatial phase $\Phi_C^1(\boldsymbol{r})$ for modulations at $\boldsymbol{Q}_C^1$. (c) Measured PDW spatial phase $\Phi_P^1(\boldsymbol{r}) - 2\pi/3$ for modulations at $\boldsymbol{Q}_P^1$. The data in (b, c) were obtained in the same FOV as in (a). (**d-f**) PDW states in AC-CDW and HC-CDW regions in $NbSe_2$ single crystals [75]. (d) Energy gap map in the AC-CDW region. (e) Energy gap map in the HC-CDW region. (f) Linecuts of $\Delta(\boldsymbol{r})$ along three trajectories indicated as the blue, yellow, green arrows in panel (d), showing the $C_3$-symmetry breaking.

### 2.5.2 $MoTe_2$

$MoTe_2$, a TMD superconductor with a superconducting transition temperature of approximately 6 K in monolayers, exhibits two-gap superconductivity and a reduction in $T_c$ due to disorder [142, 153]. Wei *et al.* reported the coexistence of a unidirectional charge order and a primary PDW in this material [76]. They observed an energy-independent charge order exhibiting particle-hole asymmetry in topographic image $T(\boldsymbol{r})$ and its Fourier transform, characterized by the wavevector $\boldsymbol{Q}_{1q-5a} = \frac{1}{5}\boldsymbol{Q}_{\text{Bragg}}$. Moreover, energy gap $\Delta(\boldsymbol{r})$, coherence peak height $H(\boldsymbol{r})$, and integrated density of states $I(\boldsymbol{r}, V)$ within the gap all exhibited periodic modulation at the same wavevector while maintaining particle-hole symmetry, suggesting the correlation with CDW and the signatures of a striped PDW phase. Based on the distinct particle-hole symmetries of PDW and charge order, along with the observation of a CDW component at $2\boldsymbol{Q}_{1q-5a}$, the authors proposed that the PDW represents a primary order. Within this scenario, the inter-band electron pairing with finite center-of-mass momentum may drive the striped PDW. Crucially, the direct observation of the striped PDW offers a clear rationale for the twofold symmetry noted in the in-plane critical field of few-layer $MoTe_2$.

Furthermore, Cheng *et al.* investigated moderately doped 1T'-$MoTe_2$ films epitaxially grown on graphitized SiC(0001) substrates and identified PDW states with wavevectors $\boldsymbol{Q}_{1\mathrm{q}-a} = \boldsymbol{Q}^{a}_{\mathrm{Bragg}}$, $\boldsymbol{Q}_{1\mathrm{q}-a/2} = 2\boldsymbol{Q}^{a}_{\mathrm{Bragg}}$, $\boldsymbol{Q}_{1\mathrm{q}-3b} = \frac{1}{3}\boldsymbol{Q}^{b}_{\mathrm{Bragg}}$ [77]. The superconducting modulations associated with $\boldsymbol{Q}_{1\mathrm{q}-a}$ and $\boldsymbol{Q}_{1\mathrm{q}-3b}$ were attributed to Amperean pairing, whereas the modulation at $\boldsymbol{Q}_{1\mathrm{q}-a/2}$ was associated with the preferential pairing of electrons between distinct Te sublattices rather than within the same sublattice. This sublattice-specific pairing gives rise to spatial modulations in $\Delta(\boldsymbol{r})$, exhibiting a half-unit-cell periodicity along the $a$-axis. Additionally, similar to observations in recent studies of cuprates, the observed pseudogap was proposed to originate from fluctuating PDWs above the superconducting transition temperature $T_c$ and the upper critical field $B_{c2}$.

### 2.6 Complementary experimental signatures and candidate materials for PDW

The superconductors discussed above have been directly observed to exhibit spatial modulations in STM experiments, but there are still many superconductors that have only been reported to exhibit PDW in other experimental methods.

The existence of PDW states in cuprates $La_{2-x}Ba_xCuO_4$ (LBCO) has been suggested by several other experimental techniques [6]. Anomalous behavior near x = 1/8, where the bulk superconducting $T_c$ is suppressed, aligns with the emergence of charge and spin stripe orders [154]. Neutron and X-ray diffraction showed stripe pinning due to lattice distortions [155], while resistivity measurements revealed a decoupling between $CuO_2$ layers, implying 2D superconductivity below the spin ordering temperature [156]. Optical reflectivity [157] and c-axis magnetic field experiments [158] reported a suppression of interlayer Josephson coupling, consistent with PDW order. Additionally, photoemission spectroscopy indicated an antinodal gap [39, 40], a signature of PDW, while Josephson current-phase relation (CPR) measurements showed a non-sinusoidal form with a second harmonic component [159], reinforcing the PDW presence. These combined results suggest that PDW order is fundamental in cuprates, leading to unique superconducting properties such as decoupling and anisotropy. Besides LBCO, various experiments on $La_{2-x}Sr_xCuO_4$ (LSCO) [160–163], Nd doped LSCO [164], $La_{2-x}Ca_{1+x}Cu_2O_6$ (LCCO) [165] have also supported the existence of PDW in cuprates.

In spin-triplet superconductors, $CeCoIn_5$ and $Sr_2RuO_4$ are expected to have PDW states as well. NMR spectra exhibited a double-horn structure, which is explained by real-space sinusoidal spin modulation, revealing a signature of the FFLO states in $Sr_2RuO_4$ [41]. A theoretical study reported that finite-momentum pairing in multiorbital systems arises from Hund's coupling and orbital non-degenerate kinetic terms. Hund's coupling provides an attractive channel in orbital-singlet spin-triplet (OSST) superconductors, enabling the stabilization of finite-momentum states, particularly under weak SOC conditions. Through analysis of a three-orbital system, they demonstrated how OSST pairing induces finite-momentum states in $Sr_2RuO_4$ and further noted that an in-plane magnetic field drives a transition from OSST pairing to the FFLO phase [166]. Many experiments have suggested the presence of PDW in $CeCoIn_5$ as reviewed in reference [167].

Beyond the materials reviewed above, nickel-based superconductors have emerged as promising candidates for detecting PDW. This potential stems from their properties analogous to those of cuprates [168], including similar crystal and electronic structures [169], strong electronic correlations [170], antiferromagnetic excitations [171, 172], and superconducting dome observed in their phase diagrams. In nickelate superconductors, the pairing symmetry is still under active debate. Some studies suggest that $d_{xy}$-wave symmetry pairing may dominate at low temperatures [171], although alternative scenarios have also been proposed [173]. Moreover, their electronic structures reveal pronounced electron correlations, particularly multi-orbital effects [174], which may facilitate the stabilization of complex electron pairing states such as PDW. Additionally, CDWs arising from complex interactions among electronic orbitals and specific lattice configurations have been observed in these materials [175], and some of them exhibit strongly intertwined charge and spin density wave states [176]. These intriguing properties suggest the existence of PDW in nickelate superconductors, but this hypothesis requires further theoretical and experimental investigations.

The observation of the temperature-dependent PDW behavior and its suppression inside magnetic vortices of $EuRbFe_4As_4$ [69] suggests a potential route to construct a PDW order from its foundational elements through the coupling of ultra-thin magnetic materials with superconductors or through forming heterostructure. When superconductors are adjacent to ferromagnetic materials, the ferromagnetic exchange field may induce spin asymmetry in Cooper pairs, potentially fostering finite-momentum pairing states and leading to PDW formation. At the interface between $KTaO_3$ and EuO, the interaction between Ta 5*d* and Eu 4*f* orbitals may drive the development of one-dimensional superconducting stripes, facilitating the emergence of PDW order [177]. These studies encourage the use of proximity effect or heterostructure to manipulate the PDW states, particularly by forming heterostructure from ferromagnetic and superconducting materials. The theoretical prediction of PDW superconductivity in TMD heterobilayers [148] inspires future experimental studies.

Recent studies utilizing the Density Matrix Renormalization Group method indicated that quantum critical points in three-dimensional Weyl semimetals and two-dimensional Dirac semimetals exhibit emergent spacetime supersymmetry within PDW superconductivity contexts [178]. Further research on two-dimensional honeycomb lattices with spinless fermions, using high-precision, unbiased quantum calculations, showed that strong interactions in these systems induce PDW superconductivity [179], enhancing the understanding of its microscopic mechanisms. In addition, these theoretical and computational results revealed that twisted moiré systems with spin polarization may be a potential candidate for detecting and manipulating PDW states [148, 180]. PDW states can also be manipulated within artificial superlattice. In some materials with strong electronic correlation effects, such as layered MBenes (two dimensional metal-boride compounds) [181] and magic-angle graphene [182], PDW states may be detected; however, further investigation is still needed.

## 3. Summary and perspective

In this topical review, we discussed recent progress in the study of PDW states, with a particular emphasis on STM/STS techniques. As a powerful local probe with unparalleled spatial resolution, STM/STS has enabled direct visualization of PDW-associated spatial modulations, such as those in the SC energy gap, condensed electron-pair density, coherence peak height, and phase-winding structure. We reviewed experimental STM/STS signatures of PDW states across a wide range of SC material systems, including cuprates, iron-based superconductors, spin-triplet superconductors, Kagome lattice superconductors, and transition metal dichalcogenides, among others. Through case-by-case analysis, we discussed how STM signatures point toward the existence of PDW states. In addition, we discussed PDM, which was proposed due to its symmetry-breaking behavior and intra-unit-cell characteristics, and which, in a broad sense, can be regarded as a type of PDW. These advances not only deepen the understanding of the nature of unconventional superconductivity but also offer valuable physical insights for future experiments and material designs.

Further exploration of PDW states in emerging material systems, along with their control and manipulation via external fields or heterostructures, and their interplay with other quantum phases, is crucial for advancing both fundamental physics and practical applications in quantum technologies. It should be noted that one of the primary challenges in detecting PDW states lies in the fact that the intricate changes in the electronic structure associated with PDW states may not produce readily distinguishable features in the LDOS as measured by STM. Therefore, ultralow temperatures STM systems with ultrahigh energy resolution combined with other phase sensitive measurements are expected to be effective in searching for PDW states in superconductors. Moreover, several questions remain unresolved, including whether the PDW constitutes a primary order or merely emerges as a secondary modulation induced by other preexisting orders, as well as the microscopic origin of PDW states and their role in the superconducting mechanism. Recent studies have further cautioned that conherence peak modulations detected by STM may alternatively originate from pair-breaking scattering interference (PBSI), rather than a genuine PDW order [183]. This alternative interpretation highlights the ongoing debates in disentagling intrinsic PDW signatures from other competing mechanisms in unconventional superconductors [184]. These ongoing debates underscore the necessity for continued experimental and theoretical investigations.

**Acknowledgements** The authors gratefully acknowledge the support from the National Natural Science Foundation of China (Grant Nos. 62488201), the CAS Project for Young Scientists in Basic Research (YSBR-003) and the Innovation Program of Quantum Science and Technology (2021ZD0302700).